\documentclass[prb,preprint]{revtex4}
\usepackage{graphicx}
\usepackage{dcolumn}
\usepackage{bm}
\usepackage{amssymb,amsmath}

\begin{document}
\title{Linking high and low temperature plasticity in bulk metallic glasses: thermal activation, extreme value statistics and kinetic freezing}
\author{P. M. Derlet}
\email{Peter.Derlet@psi.ch} 
\affiliation{Condensed Matter Theory Group, Paul Scherrer Institut, CH-5232 Villigen PSI, Switzerland}
\author{R. Maa{\ss}}
\email{robert.maass@ingenieur.de}
\affiliation{Institute f\"{u}r Materialphysik, University of G\"{o}ttingen, Friedrich-Hund-Platz 1, D-37077 G\"{o}ttingen, Germany}
\date{\today}

\begin{abstract}
At temperatures well below their glass transition, the deformation properties of bulk metallic glasses are characterised by a sharp transition from elasticity to plasticity, a reproducible yield stress, and an approximately linear decrease of this stress with increasing temperature. In the present work it shown that when the well known properties of the under-cooled liquid regime, in terms of the underlying potential energy landscape, are assumed to be also valid at low temperature, a simple thermal activation model is able to reproduce the observed onset of macro-scopic yield. At these temperatures, the thermal accessibility of the complex potential energy landscape is drastically reduced, and the statistics of extreme value and the phenomenon of kinetic freezing become important, affecting the spatial heterogeneity of the irreversible structural transitions mediating the elastic-to-plastic transition. As the temperature increases and approaches the glass transition temperature, the theory is able to smoothly transit to the high temperature deformation regime where plasticity is known to be well described by thermally activated viscoplastic models.
\end{abstract}
\maketitle

\section{Introduction}

The low temperature mechanical properties of structural glasses, such as Bulk Metallic 	Glasses (BMGs), are characterised by a large elastic strain and a well defined yield point~\cite{Schuh2007}. This latter feature is particularly the case for tensile deformation, but work has also shown that if a pure uni-axial stress state can be achieved in compression, the yield point is also reproducible with a high Weibull modulus comparable to that of a work-hardened crystalline metal~\cite{Wu2008}. Thus, although BMGs lack good ductility, they are not brittle in the ceramic sense since the relatively low Weibull modulus of a ceramic indicates a failure mechanism sensitive to macroscopic flaws. Like work hardened crystalline metals, this suggests that the structure of a BMG is relatively homogeneous above a certain length scale, and yield is an intrinsic material property. Other related properties characterising the low temperature deformation regime of BMGs are a strong spatially heterogeneous plasticity, a weak strain rate sensitivity of the yield stress and its approximately linear decrease with respect to increasing temperature.

There has been much work attempting to characterise the underlying microscopic structural mechanisms mediating plastic evolution in BMGs. In the seminal work of Spaepen~\cite{Spaepen1977}, the picture of single atom migration into regions of excess free volume was proposed. In subsequent theoretical work by Argon~\cite{Argon1979}, the atomic scale structural transformation was hypothesised to consist of a rearrangement of atoms which mainly resulted in a local shear stress relief. The early atomistic work of Falk and Langer~\cite{Falk1998} supported this picture introducing the concept and terminology of the Shear Transformation Zone (STZ), in which groups of atoms collectively undergo a local shear transformation. Static and ultra-high strain rate dynamic atomistic simulations have given much insight into the concept of the STZ --- how it may be realised at the atomic level and how together they might collectively behave for the emergence of material yield and subsequent plastic flow~\cite{Falk1998,Schuh2003,Maloney2004,Demkowicz2005,Shi2006,Rodney2009a,Rodney2009b,Rodney2011}. Indeed, recent molecular dynamics simulation work by Guan {\it et al.} \cite{Guan2010} suggests that at high enough strain rates macroscopic yield corresponds to a stress-induced reduction of the glass transition temperature, providing a link to the quite general phenomenon of jamming/unjamming~\cite{Liu1998,Trappe2001}. 

In both the works of Spaepen~\cite{Spaepen1977} and Argon~\cite{Argon1979}, the theory of thermal activation played a central role, an approach that forms the basis of many numerical mesoscopic models of quasi-static plastic flow in BMGs \cite{Bulatov1994a,Bulatov1994b,Bulatov1994c,Homer2009,Homer2010}. Indeed for temperatures outside the low temperature deformation regime, and close to the glass transition, simple thermally activated viscoplastic models can describe well the experimentally observed strong temperature and strain rate dependent deformation behaviour~\cite{Schuh2007,Heggen2005,Wang2011} --- a regime of deformation that is now spatially homogeneous and closely related to the viscosity of the under-cooled liquid regime. 

In the under-cooled liquid regime, two distinct relaxation timescales are believed to determine the physics of viscosity. These are the slow $\alpha$-relaxation processes and the fast $\beta$-relaxation processes, both of which were first postulated by considering the underlying potential energy landscape (PEL)~\cite{Goldstein1969}, and latter confirmed by experiment~\cite{Johari1970a,Johari1970b,Johari1973}.

The PEL viewpoint has afforded great physical insight into the origin of the strong temperature dependence of viscosity in approaching the glass transition temperature from the under-cooled liquid regime~\cite{Stillinger1995,Debenedetti2001}, as evidenced by the developed understanding of the well known Angell plot that introduced the important material parameter of fragility~\cite{Angell1991}. The PEL for an under-cooled liquid is seen as a complex energy landscape of hills and valleys consisting of mega-basins, whose exit characterises the $\alpha$-relaxation, and whose underlying finer structure characterises the $\beta$-relaxation~\cite{Goldstein1969,Stillinger1995,Debenedetti2001,Heuer2008}. It is in this context that $\beta$-relaxation is said to mediate $\alpha$-relaxation, a viewpoint that is also extended to temperatures below, but close to, the glass transition. In this high temperature regime (of a BMG), $\beta$-relaxation is seen as being equivalent to STZ activity causing a build up of internal stress that ultimately underlies the $\alpha$-relaxation energy landscape~\cite{Harmon2007}. Indeed, for a wide range of BMGs, there exists a strong correlation between the apparent barrier energy corresponding to $\alpha$-relaxation (derived from the experimental kinetic fragility) and the characteristic barrier energy associated with thermally activated plastic flow (derived from temperature dependent strain rate deformation behaviour)~\cite{Wang2011}.

As mentioned in the introductory paragraph, at temperatures well below the glass transition, distinctly different deformation behaviour is observed. Because of the experimentally observed weak temperature and strain rate dependence of the yield (and maximum flow) stress, there has been some focus on athermal explanations~\cite{Falk1998,Bouchbinder2007,Falk2011} in which an effective temperature is introduced arising from the inherent quenched structural disorder of the amorphous solid. However, there are experimental indications that a thermal activation mechanism also underlies the strongly inhomogeneous plasticity characteristic of this low temperature regime. This is particularly the case for compression geometries in which the initiation and propagation of shear bands and serrated versus non-serrated flow has been studied~\cite{Kimura1980,Kimura1982,Kimura1983,Dubach2009,Klaumunzer2010,Maass2011,Maass2012}. Thermal activation theories have also been used to describe the weak reduction in yield stress with increasing temperature. Using elasticity theory, Argon~\cite{Argon1979} obtained a stress dependent critical barrier energy whose leading order term resulted in the reduction of yield stress following a $T^{1/2}$ power-law~\cite{Schuh2007}. More recently, Johnson and Samwer~\cite{Johnson2005} have used a thermal activation model to derive an expression for the characteristic stress of plastic flow, which they argued is representative of the characteristic yield stress, to obtain a $(T/T_{\mathrm{g}})^{2/3}$ temperature dependence that describes well the universal trend seen in BMGs between yield stress (divided by a representative elastic modulus) and temperature. Again, the concept of a critical STZ barrier energy was used.

In this paper, the properties of the under-cooled liquid PEL are further exploited to develop a theory of thermally activated plasticity for BMGs which is valid at both high and low temperatures below the glass transition. The current work extends on previous work~\cite{Derlet2011} and gives more detail to that presented in ref.~\cite{Derlet2012}. Specifically, the complexity of the underlying PEL, in terms of the number of underlying available irreversible structural transformations and their energy distribution, is extended to temperatures below the glass transition and used to postulate a coarse grained distribution of $\alpha$-relaxation barrier energies. When used in conjunction with the thermal activation hypothesis, a plastic transition rate is obtained that characterises a temperature and stress scale at which non-negligible plasticity occurs giving a measure of the yield stress and a definition to the critical barrier energy alluded to by Argon~\cite{Argon1979} and Johnson and Samwer~\cite{Johnson2005}. Two distinct temperature regimes emerge, a high temperature regime close to the glass transition temperature in which the yield stress rises rapidly and non-linearly from zero with decreasing temperature, and a low temperature regime in which the plastic transition stress increases approximately linearly with decreasing temperature. It is found that this latter deformation regime is strongly affected by the universal phenomena of freezing and extreme value statistics of the PEL, leading naturally to an origin for the experimentally observed universal temperature dependent deformation behaviour of amorphous solids at low temperature.

\section{Model Development} \label{SecMD}

\subsection{Definitions and the development of a statistical framework} \label{SecTPZL}

To study the deformation properties of a structural glass, the characteristic time scale, $\tau_{\mathrm{p}}$, associated with plastic activity at a particular length-scale is considered. For temperatures close to and above the glass transition temperature, the viscosity is proportional to such a time scale~\cite{Debenedetti2001,Heuer2008}. At temperatures below the glass transition temperature, where the initial plasticity is characterised by a transition from elasticity to a plastic flow regime, the inverse $\left[\tau_{\mathrm{p}}\right]^{-1}$ can be viewed as a plastic rate. Such a basic description of low temperature plasticity cannot include elastic/plastic accommodation issues and any corresponding structural evolution, and is therefore unable to give quantitative insight into the nature of subsequent macroscopic plastic flow such as shear banding under compression and brittle fracture under tension. However, as a quantitative measure of the stress at which significant plasticity begins to occur, such a description will suffice and indeed has been used in the past to describe the transient yield strength prior to macro-plasticity~\cite{Argon1979,Schuh2007} and the yield rate by Johnson and Samwer~\cite{Johnson2005}.

If there exists a certain length scale above which a glass maybe considered homogeneous, then all material volume elements of a size comparable to this length scale are equivalent. It is the calculation of $\left[\tau_{\mathrm{p,RVE}}\right]^{-1}$ for such a representative volume element (RVE) that is of current interest. A RVE has well defined properties because a sufficient level of self averaging occurs with respect to a shorter length-scale which characterises the underlying heterogeneity. Let the volume of such a heterogeneity length-scale contain a characteristic number of atoms, $N$. It is with respect to this heterogeneous volume element that the self averaging is performed, giving
\begin{equation}
\left[\tau_{\mathrm{p,RVE}}\right]^{-1}(T)=\sum_{n=1}^{N'}\left[\tau_{\mathrm{p},n}\right]^{-1}(T)=N'\times\left[\tau_{\mathrm{p}}\right]^{-1}(T), \label{EqnYieldRate0a}
\end{equation}
where $N'$ is a sufficiently large number of heterogeneous volume elements such that the average
\begin{equation}
\left[\tau_{\mathrm{p}}\right]^{-1}(T)=\frac{1}{N'}\sum_{n=1}^{N'}\left[\tau_{\mathrm{p},n}\right]^{-1}(T), \label{EqnAve} 
\end{equation}
has converged. Thus the RVE contains $N'\times N$ atoms. In eqns.~\ref{EqnYieldRate0a} and \ref{EqnAve}, $\left[\tau_{\mathrm{p,n}}\right]^{-1}(T)$ represents the plastic rate of the $n$th heterogeneous volume element within the RVE. In the remainder of this paper it is $\left[\tau_{\mathrm{p}}\right]^{-1}(T)$ that will be considered further.

Under the assumption of thermally activated plasticity, the plastic rate associated with one particular realisation of a heterogeneity volume may be written as a linear sum of the $M=M(N)$ individual irreversible transition rates available to that volume element:
\begin{equation}
\left[\tau_{\mathrm{p,n}}\right]^{-1}(T)=\sum_{i=1}^{M}\left[\tau_{\mathrm{p0},ni}\right]^{-1}(T)\exp\left(-\frac{E_{\mathrm{p0},ni}}{k_{\mathrm{B}}T}\right), \label{EqnYieldRate0b}
\end{equation}
where $\left[\tau_{\mathrm{p0},ni}\right]^{-1}(T)$ and $E_{\mathrm{p0},ni}$ are the attempt rate and barrier energy for the $i$th irreversible structural transformation within the $n$th heterogeneous volume element, respectively. Substitution of eqn.~\ref{EqnYieldRate0b} into \ref{EqnAve} then gives
\begin{equation}
\left[\tau_{\mathrm{p}}\right]^{-1}(T)=M\left\langle\left[\tau_{\mathrm{p0}}\right]^{-1}(T)\exp\left(-\frac{E_{\mathrm{p0}}}{k_{\mathrm{B}}T}\right)\right\rangle, \label{EqnYieldRate}
\end{equation}
where $\langle\cdots\rangle$ represents an average with respect to sampled realisations of heterogeneous volume elements. Such an average will be performed under the assumption that the heterogeneous volume elements are statistically independent from each other.

\begin{figure}
\begin{center}
\includegraphics[clip,width=0.6\textwidth]{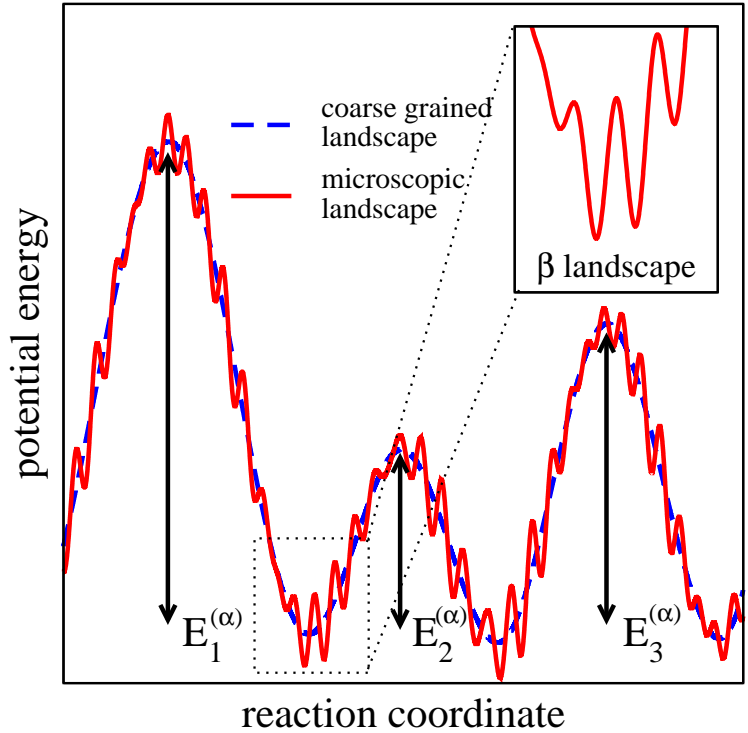}
\end{center}
\caption{Schematic of a one-dimensional realisation of the potential energy landscape of a structural glass. The horizontal reaction coordinate represents a trajectory in $3N$ dimensional space in which the system moves from minimum to minimum via a saddle-point configuration. The red-line represents the actual microscopic landscape and demonstrates that many intermediate minima ($\beta$-relaxation/STZ processes) must be traversed before exiting a mega-basin ($\alpha$-relaxation process). The blue dashed line indicates a coarse graining of the landscape, in which only the $\alpha$-relaxation landscape is retained.} \label{FigPEL} 
\end{figure}

To further simplify eqn.~\ref{EqnYieldRate}, knowledge of the underlying PEL of an amorphous solid is required. Fig.~\ref{FigPEL} displays a typical one dimensional schematic of a microscopic PEL indicating the two characteristic energy landscape scales of the $\alpha$-relaxation and $\beta$-relaxation processes. This figure schematically shows that to escape the mega-basins associated with the $\alpha$-relaxation processes, multiple (and reversible) $\beta$-relaxation (STZ) activity will occur~\cite{Heuer2008}. Exiting such mega-basins results in irreversible structural transformations and thus emergent meso/macro-scopic plasticity. It is such structural transformations which enter explicitly into eqn.~\ref{EqnYieldRate0b}, and thus the barrier energies $E_{\mathrm{p0},ni}$ will represent those of the available $\alpha$-relaxation processes (for example, the barrier energies $E_{1}^{(\alpha)}$, $E_{2}^{(\alpha)}$ and $E_{3}^{(\alpha)}$ schematically shown in fig.~\ref{FigPEL}), whereas the multiple mediating $\beta$-relaxation/STZ activity will enter implicitly into the temperature dependent prefactors of eqn.~\ref{EqnYieldRate0b}. This separation of energy scales and dynamics represents a natural coarse graining of the PEL in which the microscopic landscape of the $\beta$-relaxation is effectively integrated out resulting in prefactors which are themselves diffusive and thus based on thermal activation.

The average in eqn.~\ref{EqnYieldRate} is with respect to the different possible temperature dependent pre-factors, $\left[\tau_{\mathrm{p0},ni}\right]^{-1}(T)$ and mega-basin ($\alpha$-relaxation) barrier energies $E_{\mathrm{p0},ni}$ and can be performed via a corresponding probability distribution with respect to these variables. Such a distribution naturally acknowledges that a BMG can admit a wide range of irreversible microscopic processes, a feature that has been discussed and exploited in past work on the early stages of plasticity~\cite{Argon1979,Argon1968,Argon1980,Khonik2000} and more structural relaxation~\cite{Gibbs1993,Khonik1998,Khonik2001}.

What form should this distribution take? Because the attempt rate for an identified escape path in the coarse grained landscape will depend on the specific geometry of the underlying microscopic landscape, its actual value and temperature dependence will be quite unique to that escape route and no one particular value is expected to be a representative case. This realisation results in the considerable simplification of allowing independent averaging of the pre-factor and the barrier energy, an approximation that has also been rationalized by Gibbs {\em et al} \cite{Gibbs1993} and whose consequences have been discussed by Primak\cite{Primak1955}. Eqn.~\ref{EqnYieldRate} then becomes 
\begin{equation}
\left[\tau_{\mathrm{p}}\right]^{-1}=M\left\langle\left[\tau_{\mathrm{p0}}\right]^{-1}(T)\right\rangle\left\langle\exp\left(-\frac{E_{\mathrm{p0}}}{k_{\mathrm{B}}T}\right)\right\rangle. \label{EqnYieldRate1}
\end{equation}

Since thermal activation of multiple $\beta$-relaxation/STZ processes mediates the $\alpha$-relaxation process, a natural choice for the average prefactor will be
\begin{equation}
\left\langle\left[\tau_{\mathrm{p0}}\right]^{-1}(T)\right\rangle=\left[\tau_{\mathrm{p00}}\right]^{-1}\exp\left(-\frac{E_{\mathrm{p00}}}{k_{\mathrm{B}}T}\right), \label{EqnYieldRate3}
\end{equation}
where $E_{\mathrm{p00}}$ is a characteristic barrier energy for $\beta$-relaxation. This form is motivated by experiment which demonstrates that the average temperature dependence of $\beta$ relaxation is Arrhenius both above and below the glass transition temperature~\cite{Johari1970b,Stillinger1995,Debenedetti2001}. The lack of a temperature dependence in the corresponding characteristic pre-factor is also motivated by the same assertion leading to the independent averaging within eqn.~\ref{EqnYieldRate}, where no particular pre-factor (and its temperature dependence) is representative --- a situation that upon averaging naturally leads to little or no systematic temperature dependence. It will turn out, that any temperature dependence that might arise from the averaging of eqn.~\ref{EqnYieldRate3} will play only a minor role in the underlying properties of the model.

That a weak correlation between prefactor and barrier energy exists within an amorphous solid, has recently been demonstrated via atomistic simulations of model Lenard-Jones glasses~\cite{Koziatek2013} using the activated-relaxation technique (ART) of Barkema and Mousseau~\cite{Barkema1996,Mousseau1998,Olsen2004}. The ART method involves exploration of the nearby PEL of a local minimum and identification of multiple saddle-point atomic configurations, thereby giving a distribution of barrier energies\cite{Rodney2009a,Rodney2009b,Kallel2010} for the $\beta$ relaxation landscape. In ref.~\cite{Koziatek2013}, the prefactor for each identified barrier energy was calculated according to harmonic transition state theory and shown to span several orders of magnitude in value and be only very weakly correlated with barrier energy. 

Thus the final equation for the average plastic rate of a heterogeneous volume is 
\begin{equation}
\left[\tau_{\mathrm{p}}\right]^{-1}=\left[\tau_{\mathrm{p00}}\right]^{-1}\exp\left(-\frac{E_{\mathrm{p00}}}{k_{\mathrm{B}}T}\right)M\left\langle\exp\left(-\frac{E_{\mathrm{p0}}}{k_{\mathrm{B}}T}\right)\right\rangle, \label{EqnYieldRate5}
\end{equation}
consisting of a diffusive attempt rate with a simple Arrhenius temperature dependence, representing the mediating $\beta$-relaxation dynamics, and a thermal factor whose temperature dependence will be derived from the statistical properties of the $\alpha$-relaxation coarse grained PEL.  

\subsection{A Gaussian distribution of barrier energies, extreme value statistics and the emergence of kinetic freezing} \label{SecGauss}

How is $M$ related to $N$ and what is the form of the barrier energy distribution, $P(E)$? 

There will exist a relation between $M$ and $N$ whose origin lies in the properties of the PEL. Stillinger has shown the number of minima (inherent structures) that a structural glass can admit scales exponentially with system size \cite{Stillinger1983,Stillinger1984,Stillinger1999}. Later work by Scott Shell {\em et al} \cite{Shell2004} extends such exponential scaling to saddle points, as was also done for a many-particle random Gaussian potential by Fyodorov \cite{Fyodorov2004}. A toy model that approximates the minima of the PEL as an $N$ hyper-dimensional cube~\cite{Kohen2000} has also been extended to the enumeration of saddle points resulting in a similar exponential scaling of the number of saddle points~\cite{Derlet2011}. Thus the relation between $M$ and $N$ is generally excepted to be exponential, $M=\exp(\overline{\alpha}N)$, where for the case of counting minima (inherent structures), $\overline{\alpha}$ is a measure of the configurational entropy per atom. Such an exponential scaling of the number of available transition paths is not new and has been assumed in thermodynamical treatments of the viscosity of an under-cooled liquid~\cite{Adam1965,Kirkpatrick1989}. In the present case, $M$ counts the number of first order saddle points in the coarse grained PEL and $\overline{\alpha}$ can be referred to as the configurational {\em barrier} entropy per atom.

The form of $P(E)$ turns out to be intimately related to the above exponential scaling, where the number density with respect to barrier energy of $\overline{\alpha}$-relaxation barrier energies may be written as $M(E)=MP(E)$. Similar to Stillinger's work on the distribution of inherent structures as a function of system energy~\cite{Debenedetti2001}, the work of Scott Shell {\em et al} \cite{Shell2004}, and Derlet and Maass \cite{Derlet2011} (within the frame work of a toy model), have shown that such a number density can be well approximated by a Gaussian distribution, giving
\begin{equation}
P(E)=\frac{1}{\sqrt{2\pi\langle E^{2}\rangle}}\exp\left[-\frac{\left(E-\langle E\rangle\right)^{2}}{2\langle E^{2}\rangle}\right], \label{EqnGD}
\end{equation}
with its first and second cumulants being extensive quantities, giving the mean as $\overline{E}N$ and the standard deviation as $\delta\overline{E}\sqrt{N}$. Here $\overline{E}$ can be viewed as the mean $\alpha$-relaxation barrier energy per atom. Whilst such a distribution will describe well the true distribution in the region where it is non negligible in value, it will give an overestimate of the barrier probability in the low barrier regime, indeed at a barrier energy equal to zero it will give a non-zero probability. Whilst this aspect will be addressed in sec~\ref{SecLoad}, further development will assume that $\overline{E}N$ and $\delta\overline{E}\sqrt{N}$ have values such that $P(E<0)$ is negligible. In any event, the defining property of the distribution is not its precise form, but rather that it has an extensive first and second cumulant. 

Inspection of eqn.~\ref{EqnYieldRate5} reveals that the plastic rate is proportional to $M\langle\exp(-\beta E)\rangle$ (here $\beta=1/(k_{\mathrm{B}}T)$) where $\langle\exp(-\beta E)\rangle$ is the generating function of the distribution evaluated at $-\beta$. For the Gaussian distribution the generating function has a closed form, resulting in 
\begin{eqnarray}
M\left\langle\exp\left(-\frac{E}{k_{\mathrm{B}}T}\right)\right\rangle&=&M\int_{0}^{\infty}\,dE\,P(E)\exp\left(-\frac{E}{k_{\mathrm{B}}T}\right) \nonumber \\ 
&\approx&\exp\left(N\left[\overline{\alpha}-\frac{\overline{E}}{k_{\mathrm{B}}T}+\frac{1}{2}\left(\frac{\delta\overline{E}}{k_{\mathrm{B}}T}\right)^{2}\right]\right). \label{EqnAveYieldRate3}
\end{eqnarray}
In the above the approximation arises due to the lower integral limit being extended from zero to $-\infty$.

\begin{figure}
\begin{center}
\includegraphics[clip,width=0.9\textwidth]{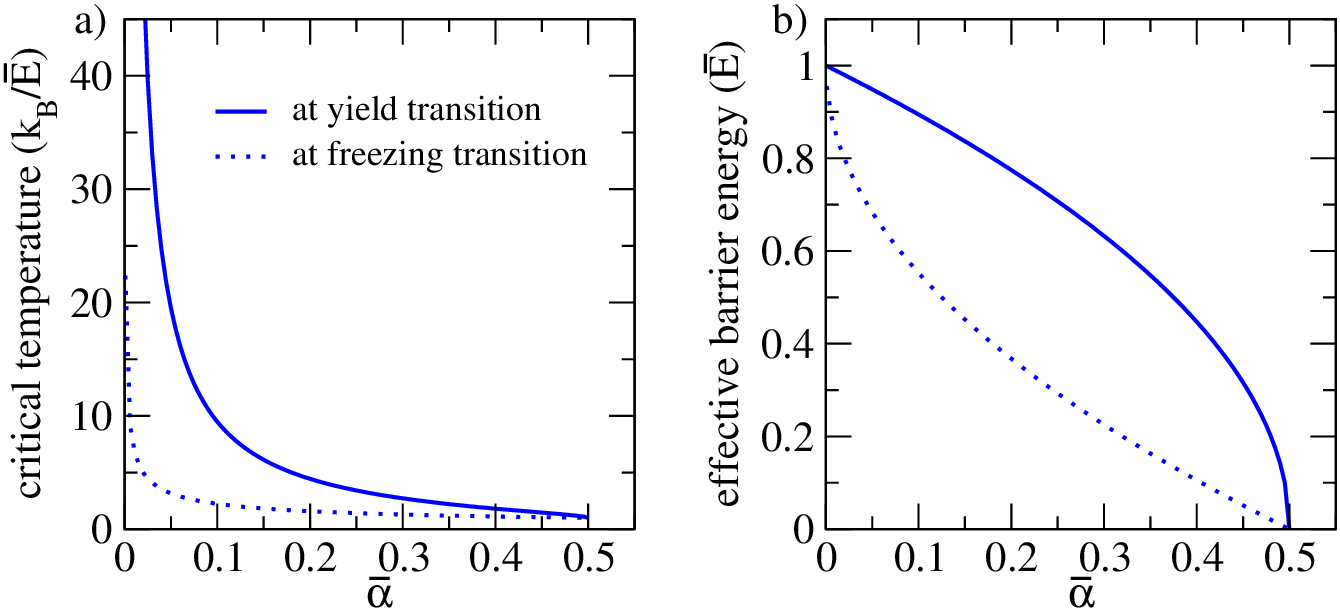}
\end{center}
\caption{Plot of a) plastic and freezing transition temperature, and b) apparent barrier energy evaluated at both the plastic and freezing transition temperatures as a function of $\overline{\alpha}$.} \label{FigCriticalTemp} 
\end{figure}

For a sufficiently large $N$, the above (and $\left[\tau_{\mathrm{p}}\right]^{-1}$ via eqn.\ref{EqnYieldRate5}) will have a negligible value when the factor within the argument of the exponential is negative and a large value when it is positive. Assuming $\overline{E}$, $\delta\overline{E}$ and $\overline{\alpha}$ are fixed material parameters, the critical temperature at which this occurs is defined by the condition
\begin{equation}
\overline{\alpha}-\frac{\overline{E}}{k_{\mathrm{B}}T_{\mathrm{c}}}+\frac{1}{2}\left(\frac{\delta\overline{E}}{k_{\mathrm{B}}T_{\mathrm{c}}}\right)^{2}=0, \label{EqnCriticalTemp0}
\end{equation}
whose relevant solution is,
\begin{equation}
T_{\mathrm{c}}=\frac{\overline{E}}{k_{\mathrm{B}}}\frac{\left(\delta\overline{E}/\overline{E}\right)^{2}}{1-\sqrt{1-2\left(\delta\overline{E}/\overline{E}\right)^{2}\overline{\alpha}}}. \label{EqnCriticalTemp1}
\end{equation}
The critical temperature is thus independent of $N$ and therefore the size of the heterogeneous volume element. $T_{\mathrm{c}}$ will be referred to as the plastic transition temperature. Fig.~\ref{FigCriticalTemp}a plots eqn.~\ref{EqnCriticalTemp1} as a function of $\overline{\alpha}$ for the special case of $\overline{E}=\delta\overline{E}$. For $\overline{\alpha}\rightarrow0$ the plastic transition temperature diverges and at its maximum allowed value of one half, $T_{\mathrm{c}}=\overline{E}/k_{\mathrm{B}}$. Thus with an increasing $\overline{\alpha}$ (an increasing number of available structural transformations), $T_{\mathrm{c}}$ reduces.

This latter result may be understood by investigating the first and second moment of the distribution, $P(E)\exp(-\beta E)$, which is straightforwardly given by the general expression for the $n$th moment,
\begin{equation}
E^{(n)}=\frac{1}{G(-\beta)}\frac{\partial^{n} G(-\beta)}{\partial(-\beta)^{n}}, \label{EqnBEMoment}
\end{equation}
where $G(-\beta)=\left\langle\exp(-\beta E)\right)\rangle$ is the generating function for the Gaussian distribution and $\beta=1/(k_{\mathrm{B}}T)$. This gives the first moment as
\begin{equation}
E^{(1)}(T)=N\left(\overline{E}-\frac{\delta\overline{E}^{2}}{k_{\mathrm{B}}T}\right), \label{EqnBEMoment1}
\end{equation}
and the second moment as
\begin{equation}
E^{(2)}(T)=N\delta\overline{E}^{2}+N^{2}\left(\overline{E}-\frac{\delta\overline{E}^{2}}{k_{\mathrm{B}}T}\right)^{2}. \label{EqnBEMoment2}
\end{equation}
The first moment will be referred to as the apparent barrier energy, $E^{(1)}(T)=N\overline{E}_{\mathrm{app}}(T)$,
\begin{equation}
\overline{E}_{\mathrm{app}}(T)=\overline{E}-\frac{\delta\overline{E}^{2}}{k_{\mathrm{B}}T}, \label{EqnAppEn}
\end{equation}
which (via eqn.~\ref{EqnBEMoment2}) has a temperature independent standard deviation equal to $\delta\overline{E}\sqrt{N}$ indicating it to be a statistically meaningful quantity for sufficiently large $N$.

At the plastic transition temperature, the apparent barrier energy becomes
\begin{equation}
\overline{E}_{\mathrm{app}}(T_{\mathrm{c}})=\overline{E}\sqrt{1-2\left(\delta\overline{E}/\overline{E}\right)^{2}\overline{\alpha}}. \label{EqnBEMoment1Critical}
\end{equation}
Eqn.~\ref{EqnBEMoment1Critical} is plotted in fig.~\ref{FigCriticalTemp}b as a function of $\overline{\alpha}$ for the special case of $\overline{E}=\delta\overline{E}$. The decrease of $\overline{E}^{\mathrm{app}}(T_{\mathrm{c}})$ with respect to $\overline{\alpha}$ reflects that $T_{\mathrm{c}}$ decreases with $\overline{\alpha}$.

As $\overline{\alpha}\rightarrow0$, the apparent barrier energy, $\overline{E}_{\mathrm{app}}(T_{\mathrm{c}})$, limits to $N\overline{E}$ which is the most probable barrier energy of $P(E)$. On the other hand, when $\overline{\alpha}$ equals one half, the apparent barrier energy becomes zero. These results can be understood by first realising that because of the thermal exponential factor in eqn.~\ref{EqnAveYieldRate3}, low barrier energies will dominate the average from which  $E_{\mathrm{app}}(T_{\mathrm{c}})$ is obtained. For example, given $M$ sampled barriers, it will be the lowest barrier energy of this list that makes the largest contribution to the apparent barrier energy. Thus the statistics of extreme values naturally emerge: sampling the distribution once (with $\overline{\alpha}=0$) results on average in the most probable value of the distribution (which is $N\overline{E}$), whereas sampling the distribution an increasing number of times will lead to lower barrier energies dominating the apparent barrier energy. The fact that $\overline{\alpha}$ is limited to being less than or equal to one half is an artifact of the Gaussian distribution not limiting to zero as the barrier energy limits to zero, giving $\overline{E}_{\mathrm{app}}(T_{\mathrm{c}})=0$ at $\overline{\alpha}=1/2$. A distribution for which $P(E\rightarrow0)\rightarrow0$ would allow for an infinite range of $\overline{\alpha}$ and a finite positive $\overline{E}_{\mathrm{app}}(T_{\mathrm{c}})$ resulting in an everlasting approach to the zero barrier energy with increasing $\overline{\alpha}$. In sec.~\ref{SecLoad}, a modification to the Gaussian distribution will be investigated that gives this correct limiting behaviour.

Eqn.~\ref{EqnBEMoment1} suggests that eqn.~\ref{EqnAveYieldRate3} may be written as 
\begin{equation}
M\left\langle\exp\left(-\frac{E}{k_{\mathrm{B}}T}\right)\right\rangle=M_{\mathrm{app}}(T)\exp\left(-\frac{N\overline{E}_{\mathrm{app}}(T)}{k_{\mathrm{B}}T}\right),\label{EqnAveYieldRate4}
\end{equation}
where $M_{\mathrm{app}}(T)$ is given by
\begin{equation}
M_{\mathrm{app}}(T)=\exp\left(N\overline{\alpha}_{\mathrm{app}}(T)\right). \label{EqnAppM}
\end{equation}
with
\begin{equation}
\overline{\alpha}_{\mathrm{app}}(T)=\overline{\alpha}-\frac{1}{2}\left(\frac{\delta\overline{E}}{k_{\mathrm{B}}T}\right)^{2}. \label{EqnAppalpha}
\end{equation}
In eqn.~\ref{EqnAveYieldRate4}, the first factor is interpreted as the apparent number of structural transformations and the second factor is associated with the apparent barrier energy of eqn.~\ref{EqnAppEn}. At high enough temperatures, the apparent energy limits to $N\overline{E}$, the most probable value of the barrier energy distribution. Thus in the high temperature limit, the entire distribution contributes to the barrier kinetics, a fact that is reflected by the apparent number of available structural transformations limiting to $M$ --- the total number of structural transformations within each heterogeneous volume element. As the temperature lowers, both the apparent barrier energy and the apparent number of available structural transformations reduce because, with a lower temperature, the higher barrier energies become increasingly unlikely to occur. Eqn.~\ref{EqnCriticalTemp0} shows that if $T>T_{\mathrm{c}}$ then eqn.~\ref{EqnAveYieldRate4} will be exponentially large because the thermal factor is over-compensated by the apparent number of structural transformations. When $T<T_{\mathrm{c}}$ this turns to an under-compensation, resulting in eqn.~\ref{EqnAveYieldRate4} becoming neglible.

Is the functional form of eqn.~\ref{EqnAveYieldRate3} and \ref{EqnAveYieldRate4} valid at all temperatures below $T_{\mathrm{c}}$? 

Inspection of eqns.~\ref{EqnAppM} and \ref{EqnAppalpha}, reveals that at a low enough temperature the argument within the exponential will equal zero resulting in the apparent number of available structural transformations equalling unity. This will occur at the temperature
\begin{equation}
T_{\mathrm{f}}=\frac{\delta\overline{E}}{k_{\mathrm{B}}}\frac{1}{\sqrt{2\overline{\alpha}}}. \label{EqnFreezingTemp}
\end{equation}
When $T<T_{\mathrm{f}}$, the apparent number of available structural transformations will be less than one. This reflects that at this temperature some heterogeneous volumes will contain no available structural transformations with a barrier energy equal to $\overline{E}_{\mathrm{app}}(T)$, and the dominant barrier energy will (on average) be the lowest available one, $\overline{E}_{\mathrm{app}}(T_{\mathrm{f}})$, which equals
\begin{equation}
\overline{E}_{\mathrm{app}}(T_{\mathrm{f}})=\overline{E}-\delta\overline{E}\sqrt{2\overline{\alpha}}. \label{EqnAppEF}
\end{equation}
Upon further decrease in temperature, $E_{\mathrm{app}}(T_{\mathrm{f}})$ will continue to dominate where now the correct expression for $T<T_{\mathrm{f}}$ is no longer eqn.~\ref{EqnAveYieldRate3} or \ref{EqnAveYieldRate4}, but rather
\begin{equation}
M\left\langle\exp\left(-\frac{E}{k_{\mathrm{B}}T}\right)\right\rangle=\exp\left(-\frac{N\overline{E}_{\mathrm{app}}(T_{\mathrm{f}})}{k_{\mathrm{B}}T}\right). \label{EqnAveYieldRateFreezing}
\end{equation}

Both $T_{\mathrm{f}}$ and $\overline{E}_{\mathrm{app}}(T_{\mathrm{f}})$ are plotted as a function of $\overline{\alpha}$ in fig.~\ref{FigCriticalTemp} showing that $T_{\mathrm{f}}$ and $\overline{E}_{\mathrm{app}}(T_{\mathrm{f}})$ generally occur below their plastic transition counterparts.

Analogous low temperature behaviour is also seen in spin-glass systems, however with respect to thermodynamic variables where at a sufficiently low temperature, $T_{\mathrm{f}}$, the thermodynamic entropy of the spin-glass becomes zero resulting in a temperature independent free energy upon further decrease in temperature. This is referred to as freezing, where at $T<T_{\mathrm{f}}$ one configuration of the system dominates the statistics giving an entropy of zero. This was first derived for the Random Energy Model by Derrida~\cite{Derrida1980} using a micro-canonical ensemble approach. Because, eqns.~\ref{EqnAveYieldRate3} and \ref{EqnAveYieldRate4} have the structure of an average partition function, the mathematical formalism associated with the micro-canonical approach of Derrida can be also used as an alternative derivation of the right-hand-side of eqn.~\ref{EqnAveYieldRate4}, and is presented in appendix~\ref{AppTherm}. Carrying out this analogy allows the apparent barrier energy to be viewed as an internal {\em barrier} energy and the natural logarithm of the apparent number of available structural transformation as a {\em barrier} entropy, resulting in eqn.~\ref{EqnAveYieldRate4} being written as
\begin{equation}
M\left\langle\exp\left(-\frac{E}{k_{\mathrm{B}}T}\right)\right\rangle=\exp\left(-\frac{F(T)}{k_{\mathrm{B}}T}\right), \label{EqnAveYieldRate5}
\end{equation}
where
\begin{equation}
F(T)=-k_{\mathrm{B}}TN\left(\overline{\alpha}-\frac{\overline{E}}{k_{\mathrm{B}}T}+\frac{1}{2}\left(\frac{\delta\overline{E}}{k_{\mathrm{B}}T}\right)^{2}\right) \label{EqnFreeEnergy}
\end{equation}
is the free {\em barrier} energy. For $T<T_{\mathrm{f}}$, $F(T)=F(T_{\mathrm{f}})$.

Within this analogy, the plastic transition temperature $T_{\mathrm{c}}$ represents the temperature at which the free {\em barrier} energy is zero --- a condition that reflects equal contributions from the internal {\em barrier} energy and the {\em barrier} entropy. In other words, $T_{\mathrm{c}}$ corresponds to the transition from an internal {\em barrier} energy dominated to a {\em barrier} entropy dominated plastic rate\cite{Derlet2011}. It must be emphasised that the internal {\em barrier} energy, {\em barrier} entropy and free {\em barrier} energy are {\em not} true thermodynamic variables, but rather represent an alternative derivation and interpretation of eqn.~\ref{EqnAveYieldRate4}. Such an analogy also gives a proper context to the meaning of $T_{\mathrm{f}}$, eqn.~\ref{EqnFreezingTemp}, which will be now referred to as the kinetic freezing temperature --- a temperature below which the barrier statistics are dominated by a single barrier energy scale.

\begin{figure}
\begin{center}
\includegraphics[clip,width=0.9\textwidth]{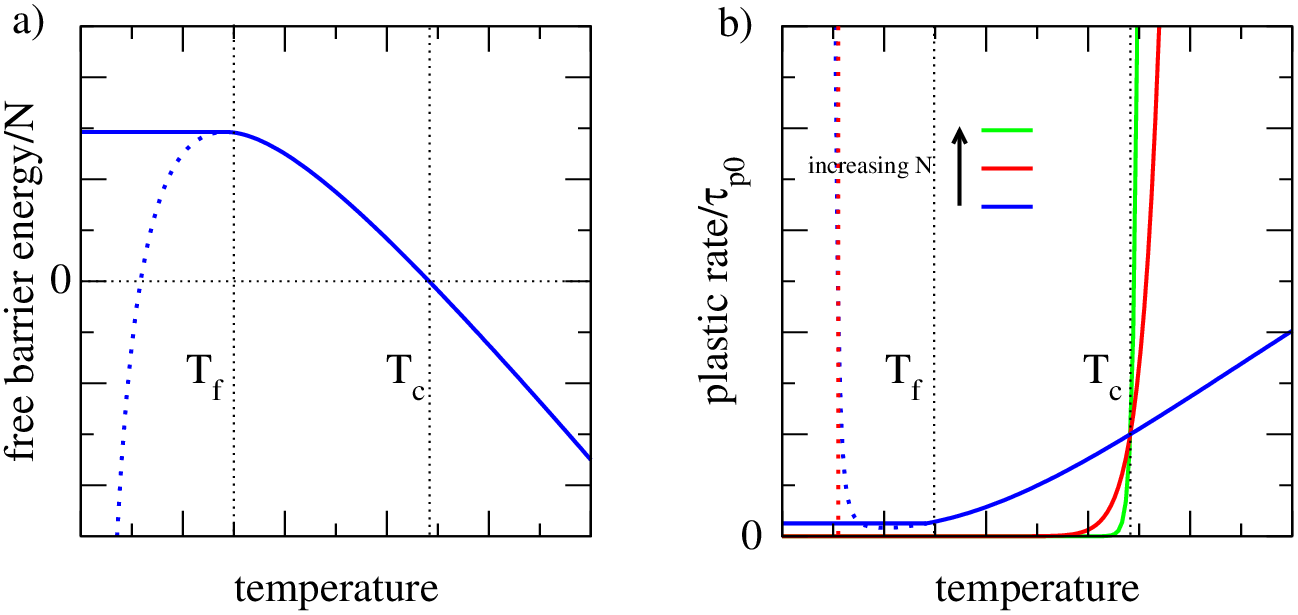}
\end{center}
\caption{Plot of a) free barrier energy and b) plastic rate as a function of temperature.} \label{FigFEYR} 
\end{figure}

Figs.~\ref{FigFEYR}a and b plot the free barrier energy, eqn.~\ref{EqnFreeEnergy}, and eqn.~\ref{EqnAveYieldRate5} (or equivalently \ref{EqnAveYieldRate4}), taking eqn.~\ref{EqnAveYieldRateFreezing} into account, as a function of temperature where both the $T_{\mathrm{f}}$ and $T_{\mathrm{c}}$ are indicated. In fig.~\ref{FigFEYR}b, the proper treatment of the low temperature behaviour becomes apparent, where with decreasing temperature eqn.~\ref{EqnAveYieldRate5} becomes a constant when $T<T_{\mathrm{f}}$. Without such a correction, decreasing the temperature below $T_{\mathrm{f}}$ would result in the unphysical increase of eqn.~\ref{EqnAveYieldRate5} and therefore the plastic rate. Fig.~\ref{FigFEYR}b also demonstrates that the sharpness of the transitions is controlled by $N$.

\subsection{The application of a load} \label{SecLoad}

The application of an external stress will affect the distribution of barrier energies via a change in its first and second moments, $\langle E\rangle$ and $\langle E^{2}\rangle$. What should the stress dependence of these moments be? For the present work, only a pure shear stress will be considered. 

The occurrence of a structural transformation will result in a localized rearrangement of atoms, which in the far-field limit can be accommodated by a homogeneous distortion of the material~\cite{Eshelby1954,Eshelby1957,Argon1979}. Thus the irreversible structural transformation will change the global stress state of the system. Since the zero-load $P(E)$ has no knowledge of an external load, any structural transformation corresponding to barrier energy $E$ could with equal chance aid or hinder the applied stress state. In a thermal activation picture of plasticity, this corresponds to the equal chance of a particular barrier energy either increasing or decreasing upon the application of the pure stress. Hence, a broadening of the distribution of barrier energies will occur. The present work therefore considers a stress independent first cumulant and a shear stress dependent second cumulant. The special case of a stress dependent first and second cumulant such that $\overline{E}=\delta\overline{E}$ is maintained was considered in ref.~\cite{Derlet2011} and results in a yield criterion that is operationally similar to that derived by Johnson and Samwer~\cite{Johnson2005}.

By symmetry, the broadening should be an even function of the pure shear stress, giving the leading order stress dependence of the second moment as
\begin{equation}
\delta\overline{E}(\sigma)=\delta\overline{E}(0)\left(1+\left(\frac{\sigma}{\sigma_{0}}\right)^{2}\right), \label{EqnLinBro}
\end{equation}
with $\delta\overline{E}(0)$ equal to what was previously referred to as $\delta\overline{E}$. When assuming a linear stress dependence of the energy barriers and a Gaussian distribution of activation volumes centered on zero, this result becomes exact. The parameter $\sigma_{0}$ will be discussed in more detail in secs.~\ref{SecAER} and \ref{SecDis}. Use of eqn.~\ref{EqnLinBro} results in the yield transition temperature, eqn.~\ref{EqnCriticalTemp1}, becoming
\begin{equation}
T_{\mathrm{c}}(\sigma)=\frac{\overline{E}}{k_{\mathrm{B}}}\frac{\left(\delta\overline{E}(0)/\overline{E}\right)^{2}\left(1+(\sigma/\sigma_{0})^{2}\right)^{2}}{1-\sqrt{1-2\left(\delta\overline{E}(0)/\overline{E}\right)^{2}\left(1+(\sigma/\sigma_{0})^{2}\right)^{2}\overline{\alpha}}}. \label{EqnCriticalTempLoad2b}
\end{equation}
and the freezing transition temperature, eqn.~\ref{EqnFreezingTemp}, becoming
\begin{equation}
T_{\mathrm{f}}(\sigma)=\frac{\delta\overline{E}(0)}{k_{\mathrm{B}}}\frac{\left(1+(\sigma/\sigma_{0})^{2}\right)}{\sqrt{2\overline{\alpha}}}. \label{EqnFreezingTempLoad1b}
\end{equation}
Thus both $T_{\mathrm{c}}$ and $T_{\mathrm{f}}$ become functions of the applied stress. 

Inspection of eqns.~\ref{EqnCriticalTempLoad2b} and \ref{EqnFreezingTempLoad1b} reveal there exists a stress, $\sigma_{\mathrm{f}}$, at which the plastic transition temperature and the freezing temperature are equal. Indeed, $T_{\mathrm{c}}(\sigma_{\mathrm{f}})=T_{\mathrm{f}}(\sigma_{\mathrm{f}})$ gives
\begin{equation}
\left(\frac{\delta\overline{E}(0)}{\overline{E}}\right)\left(1+\left(\frac{\sigma_{\mathrm{f}}}{\sigma_{0}}\right)^{2}\right)=\frac{1}{\sqrt{2\overline{\alpha}}} \label{EqnMaxStressBroadening}
\end{equation}
and thus $k_{\mathrm{B}}T_{\mathrm{c}}(\sigma_{\mathrm{f}})=k_{\mathrm{B}}T_{\mathrm{f}}(\sigma_{\mathrm{f}})=\overline{E}/2\overline{\alpha}$, with the apparent barrier energy at freezing equal to zero (eqn.~\ref{EqnAppEF}).

\begin{figure}
\begin{center}
\includegraphics[clip,width=0.9\textwidth]{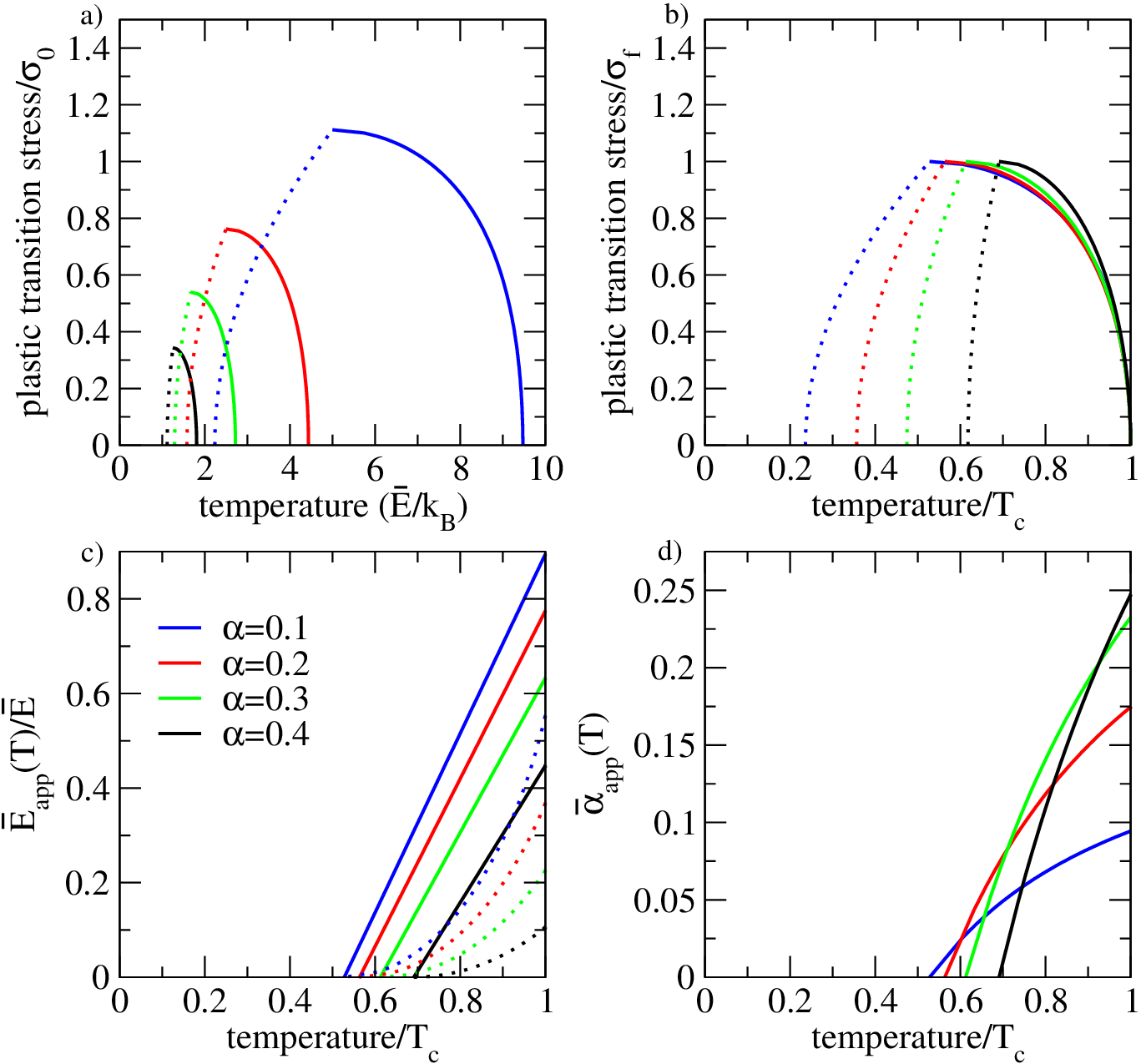}
\end{center}
\caption{a) Plot of plastic transition stress as a function temperature for values of $\overline{\alpha}$ ranging between 0.1 and 0.4. The dashed lines represent the corresponding stress dependent freezing temperature. b) Shows the same curves scaled with respect to the kinetic freezing stress, $\sigma_{\mathrm{f}}$ and the plastic transition temperature at zero-load, $T_{\mathrm{f}}$. c) Shows the apparent and freezing (dashed lines) barrier energy and d) the apparent $\overline{\alpha}$ which gives information on the number of thermally accessible structural transformations. Both c) and d) are with respect to temperature scaled with the plastic transition temperature at zero-load, $T_{\mathrm{f}}$.} \label{FigYieldTemp} 
\end{figure}

Fig.~\ref{FigYieldTemp}a plots the applied shear stress, $\sigma$ in units of $\sigma_{0}$, versus the plastic transition temperature $T_{\mathrm{c}}(\sigma)$ (eqn.~\ref{EqnCriticalTempLoad2b}) and the freezing temperature $T_{\mathrm{f}}(\sigma)$ (eqn.~\ref{EqnFreezingTempLoad1b}) for a range of $\overline{\alpha}$. In fig.~\ref{FigYieldTemp} the special case of $\overline{E}=\delta\overline{E}(0)$ is considered. Fig.~\ref{FigYieldTemp}a shows that by increasing the applied shear stress, $T_{\mathrm{c}}(\sigma)$ reduces from its zero load value, whereas $T_{\mathrm{f}}(\sigma)$ increases, eventually resulting in $T_{\mathrm{c}}(\sigma_{\mathrm{f}})=T_{\mathrm{f}}(\sigma_{\mathrm{f}})$ at the kinetic freezing shear stress $\sigma_{\mathrm{f}}$. Below this temperature eqn.~\ref{EqnCriticalTempLoad2b} is no longer applicable and the plastic transition rate becomes a constant given by eqn.~\ref{EqnAveYieldRateFreezing}. Upon increasing $\overline{\alpha}$ both the maximum yield stress and zero-load plastic transition temperature reduce. When plotted as scaled quantities, fig.~\ref{FigYieldTemp}b, the plastic transition temperature curves close to $T/T_{\mathrm{c}}$ depend relatively weakly on the choice of $\overline{\alpha}$. Fig.~\ref{FigYieldTemp}c shows the apparent barrier energy and kinetic freezing barrier energy at the plastic transition stress as a function of the plastic transition temperature. It is seen that the apparent barrier energy is at a maximum at zero load and then reduces in an approximately linear fashion with reducing temperature, finally reaching zero at freezing. The freezing barrier energy also reduces (non-linearly) to zero at freezing. Fig.~\ref{FigYieldTemp}d plots $\overline{\alpha}_{\mathrm{app}}(T)$ as a function of the plastic transition temperature, showing a non-linear reduction with respect to decreasing temperature, to zero at freezing, demonstrating that the apparent number of available structural transformations rapidly reduces to zero as the temperature decreases from $T_{\mathrm{c}}$.

The picture which therefore emerges, is that at zero load, there exists a temperature where a sufficient number of transitions exist to offset the very small thermal factor corresponding to the apparent barrier energy at that temperature. This gives a non-negligible plastic rate. At a lower temperature, upon the application of a high enough applied shear stress, such a non-negligible plastic rate can again be achieved. However, since the temperature is lower, both $\overline{E}_{\mathrm{app}}(T_{\mathrm{c}})$ and $\overline{\alpha}_{\mathrm{app}}(T_{\mathrm{c}})$ have reduced due to the higher barrier energies of the distribution becoming thermally inaccessible. Due to this reduction in the available number of structural transformations, there exists a temperature $T_{\mathrm{c}}(\sigma_{\mathrm{f}})=T_{\mathrm{f}}(\sigma_{\mathrm{f}})$ below which the system kinetically freezes before the plastic transition is obtained, irrespective of the magnitude of the applied shear stress. By increasing $\overline{\alpha}$ (and therefore increasing the total number structural transformations), both the temperature and stress at which non-negligible plasticity and/or kinetic freezing is achieved, is reduced.

Thus, with the use of a Gaussian distribution, some features of the high temperature regime of deformation of BMGs are qualitatively obtained. Fig.~\ref{FigYieldTemp}b shows that the plastic transition stress rises rapidly from zero when the temperature is decreased from the zero-load plastic transition temperature, and then saturates at a transition temperature coinciding with the applicability limit of the Gaussian model. In what follows it will be shown that, when a more realistic distribution of barrier energies is used, the developed theory may be extended to lower temperatures with a corresponding change in deformation behaviour that may be associated with the experimentally observed low temperature deformation regime of BMGs.

\subsection{A more realistic distribution of barrier energies} \label{SecGaussM}

That the freezing regime is entered in fig.~\ref{FigYieldTemp} is a direct result of the Gaussian distribution having a finite (but small) probability at the zero barrier energy. Because of this, both the apparent barrier energy and freezing barrier energy become stuck at a zero barrier energy. To modify the distribution in a way that has the correct limit $P(E\rightarrow0)\rightarrow0$ and still facilitates the developed mathematical formalism, the argument of the exponential of the Gaussian distribution must be modified. 

Starting from
\begin{equation}
P(E)\sim\exp\left[-\frac{N}{2}\left(\frac{E/N-\overline{E}}{\delta\overline{E}}\right)^{2}\right],
\end{equation}
such a modification would entail
\begin{equation}
P(E)\sim\exp\left[-\frac{N}{2}\left(\frac{g(E)/N-\overline{E}}{\delta\overline{E}}\right)^{2}\right], \label{EqnGaussM}
\end{equation}
where the function $g(E)$ is chosen such that the resulting apparent and freezing barrier energies remain extensive quantities. Additionally, $g(E\rightarrow0)\rightarrow\infty$ so that $P(E\rightarrow0)\rightarrow0$, and at larger values of $E$, $g(E)\simeq E$ to retain the Gaussian form when $E$ is comparable to $\overline{E}$. A simple choice that satisfies these requirements is
\begin{equation}
g(E)=E-\frac{\left(Na\right)^{2}}{E},
\end{equation}
where $a$ is a parameter with units of energy.

Insight into the consequences of such a modification and into the meaning of $a$ may be obtained by writing the resulting stress dependent apparent barrier energy at freezing as (see appendix~\ref{AppTherm}, eqn.~\ref{EqnFreezingBarrierApp})
\begin{equation}
\overline{E}_{\mathrm{f}}(\sigma)=\frac{\overline{E}^{\mathrm{G}}_{\mathrm{f}}(\sigma)+\sqrt{\left(\overline{E}^{\mathrm{G}}_{\mathrm{f}}(\sigma)\right)^{2}+4a^2}}{2}.
\end{equation}
Here $\overline{E}^{\mathrm{G}}_{\mathrm{f}}(\sigma)$ is the stress dependent barrier energy at freezing when using a pure Gaussian distribution, with $\overline{E}^{\mathrm{G}}_{\mathrm{f}}(\sigma_{\mathrm{f}}^{\mathrm{G}})=0$. The above equation reveals that with the modified Gaussian, the barrier energy at freezing will never become zero and at the stress, $\sigma_{\mathrm{f}}^{\mathrm{G}}$, at which freezing occurs for the pure Gaussian model, $\overline{E}_{\mathrm{f}}(\sigma_{\mathrm{f}}^{\mathrm{G}})=a$. Thus the parameter $a$ sets the apparent barrier energy at the freezing stress originally given by the Gaussian model. 

To evaluate the integral of eqn.~\ref{EqnAveYieldRate3} for this modified Gaussian, the procedure outlined in appendix~\ref{AppTherm} is used to construct the so-called free barrier energy, $F(T)$ from which, $F(T_{\mathrm{c}})=0$ can be numerically solved to obtain the critical temperature at which yield occurs for zero load conditions. 

\begin{figure}
\begin{center}
\includegraphics[clip,width=0.9\textwidth]{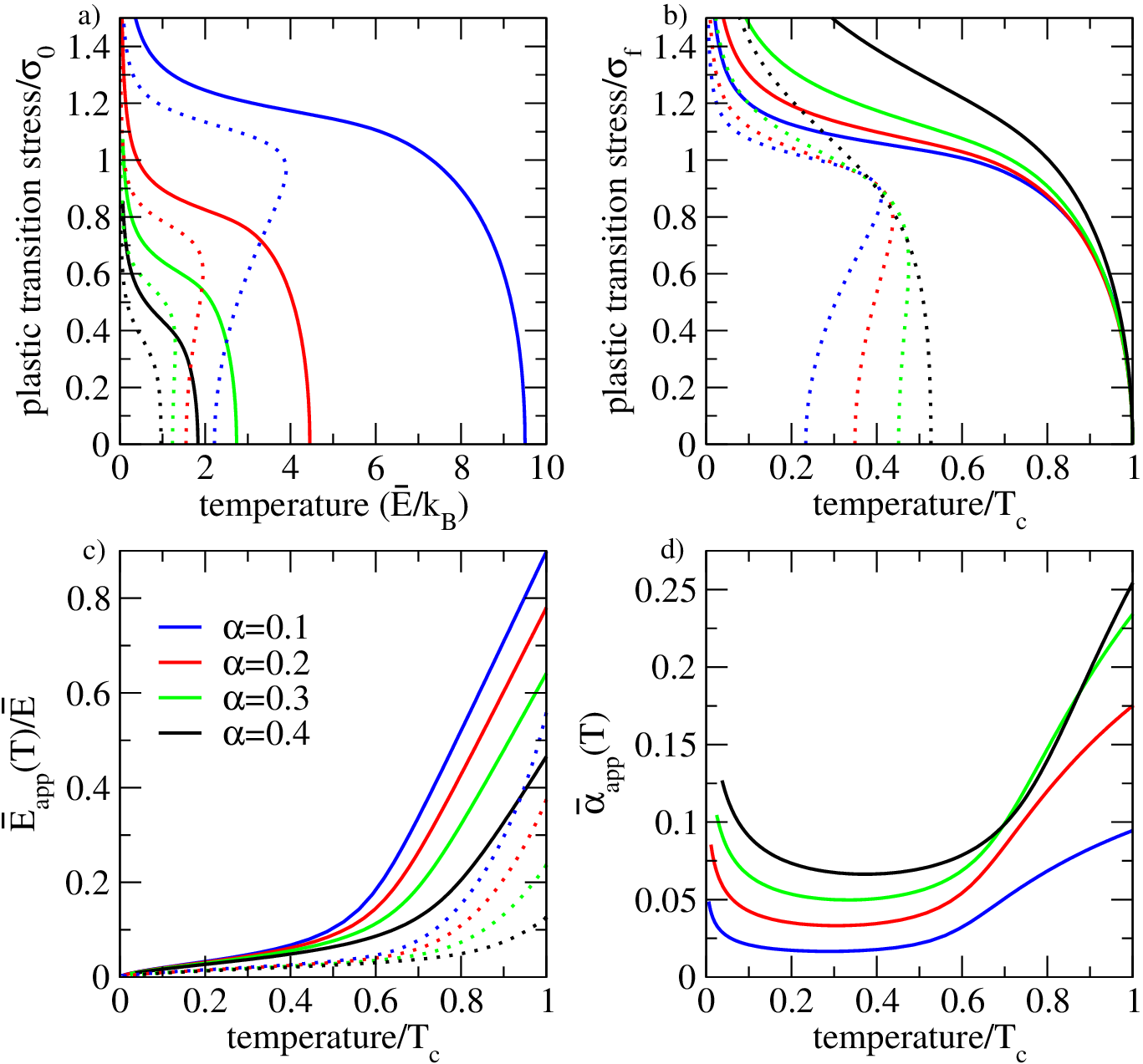}
\end{center}
\caption{Similar data to that of fig.~\ref{FigYieldTemp} using the modified Gaussian distribution of barrier energies where both a high and low temperature regime of plasticity become apparent.} \label{FigYieldTempMod} 
\end{figure}

Fig.~\ref{FigYieldTempMod} displays similar data as that shown in fig.~\ref{FigYieldTemp}. In this figure a value of $a=0.05\overline{E}$ was used. Inspection of fig.~\ref{FigYieldTempMod} reveals that at a given temperature, kinetic freezing will always occur at a stress lower than that of the plastic transition. Thus a plastic transition temperature will exist for all temperatures below the zero load plastic transition temperature. Moreover, use of the modified Gaussian distribution gives three distinct temperature regimes of behaviour: 1) a high temperature regime, in which the stress at which the plastic transition temperature occurs rises rapidly and is well described by the pure Gaussian (see fig.~\ref{FigYieldTemp} of sec.~\ref{SecLoad}), 2) a low temperature regime in which the plastic transition stress rises much more weakly (approximately linearly) with decreasing temperature and 3) an ultra-low temperature regime in which the plastic transition stress diverges. Regime 3) occurs at low enough temperatures and high enough plastic transition stresses, such that the thermal activation hypothesis becomes questionable and where an athermal regime of plasticity would emerge, a phenomenon that is beyond the scope of the present work. For the low temperature regime, 2), figs.~\ref{FigYieldTempMod}c and d, demonstrate that the modified Gaussian averts a zero apparent barrier energy and $\overline{\alpha}_{\mathrm{app}}(T)$ at finite temperature, giving a distinctly different temperature dependence to that of the high temperature regime. Decreasing $a$ is found to reduce the gradient of the approximately linear temperature dependence of the plastic transition stress, where in the limit of small $a$ a plateau is reached for temperatures less than the pure Gaussian derived $T_{\mathrm{c}}(\sigma_{\mathrm{f}})=T_{\mathrm{f}}(\sigma_{\mathrm{f}})$ value.

In summary, with the modification, $P(E\rightarrow0)\rightarrow0$, the present thermal activation model is now able to capture two distinct temperature regimes of deformation --- with increasing temperature the low temperature regime exhibits a weak decrease of the plastic transition stress, with a smooth and analytical transition to the high temperature regime of deformation where the plastic transition stress drops rapidly to zero at a finite temperature. In the next section, it will be demonstrated that the developed theory can quantitatively reproduce the corresponding experimental trends of some well known BMG materials.

\section{Application to the experimental regime} \label{SecAER}

In secs.~\ref{SecGauss} to \ref{SecGaussM}, the plastic transition stress was derived from the condition $M\left\langle\exp(-E/(k_{\mathrm{B}}T))\right\rangle=1$. For application to a real glass material, the full equation for the characteristic plastic rate (eqn.\ref{EqnYieldRate5}) must be used. The plastic transition stress must also take into account the timescale, $\tau_{\mathrm{exp}}$, of the deformation experiment. In considering the experimental timescale, a more appropriate definition of non-negligible flow to occur would be when the timescale associated with plastic activity is equal to $\tau_{\mathrm{exp}}$,
\begin{equation}
\tau_{\mathrm{exp}}\times\left[\tau_{\mathrm{p}}\right]^{-1}(T_{\mathrm{c}})\simeq1,\label{EqnYieldCondMod}
\end{equation}
where now eqn.~\ref{EqnYieldRate5} is used. Doing so, gives 
\begin{equation}
\frac{1}{N}\left[\ln\left(\frac{\tau_{\mathrm{p00}}}{\tau_{\mathrm{exp}}}\right)-\frac{E_{p00}}{k_{\mathrm{B}}T_{\mathrm{c}}}\right]+\overline{\alpha}-\frac{\overline{E}}{k_{\mathrm{B}}T_{\mathrm{c}}}+\frac{1}{2}\left(\frac{\delta\overline{E}}{k_{\mathrm{B}}T_{\mathrm{c}}}\right)^{2}=0 \label{EqnCriticalTempMod0}
\end{equation}
or
\begin{equation}
\overline{\alpha}'-\frac{\overline{E}'}{k_{\mathrm{B}}T_{\mathrm{c}}}+\frac{1}{2}\left(\frac{\delta\overline{E}}{k_{\mathrm{B}}T_{\mathrm{c}}}\right)^{2}=0, \label{EqnCriticalTempMod1}
\end{equation}
with
\begin{equation}
\overline{\alpha}'=\overline{\alpha}+\frac{1}{N}\ln\left(\frac{\tau_{\mathrm{Exp}}}{\tau_{\mathrm{p00}}}\right) \label{EqnRenormAlpha}
\end{equation}
and
\begin{equation}
\overline{E}'=\overline{E}+\frac{E_{\mathrm{p00}}}{N}. \label{EqnRenormMu}
\end{equation}
Therefore the inclusion of an experimental timescale and the prefactor, $\tau_{\mathrm{p0}}(T)$, into the plastic transition condition renormalizes the parameter $\overline{\alpha}$ and the mean barrier energy $\overline{E}$. For kinetic freezing, the corresponding bare parameters should be used.

To apply the present model to a real material, it is recognised from figs.~\ref{FigYieldTemp} and \ref{FigYieldTempMod} that the high temperature regime is well described by a pure Gaussian barrier energy distribution. Since simple analytical formulae for all relevant quantities are possible for a pure Gaussian, the high temperature deformation properties will be used to give estimates of all but one model parameter. The model parameters to consider are
\begin{enumerate}
\item $\overline{\alpha}$, the configurational barrier entropy which gives the total number of barriers per heterogeneous volume.
\item $N$, the number of atoms per heterogeneous volume.
\item $\overline{E}$ and $\delta\overline{E}$, the mean and standard deviation per atom of the barrier energy distribution.
\item the logarithm of $\tau_{\mathrm{exp}}/\tau_{\mathrm{p00}}$ and $E_{\mathrm{p00}}$.
\item $\sigma_{0}$, the shear stress sensitivity of the distribution broadening.
\end{enumerate}
Apart from $\tau_{\mathrm{exp}}$, all of the above parameters are related to the material properties of the particular structural glass of interest.

Inspection of figs.~\ref{FigYieldTemp} and \ref{FigYieldTempMod} reveal three quantities which define the range of the high temperature regime: the plastic transition temperature at zero load, and the temperature and stress at which the low temperature regime is entered (for the pure Gaussian when $T_{\mathrm{c}}(\sigma_{\mathrm{f}})=T_{\mathrm{f}}(\sigma_{\mathrm{f}})$). These three quantities may be determined directly from experiment. Indeed for a sufficiently long enough experimental timescale, $\tau_{\mathrm{exp}}$, the plastic transition temperature at zero load will be close to the material's glass transition temperature. This fact will be exploited in what follows.

The temperature at which the material enters the low temperature regime of plasticity is also experimentally well defined, being characterised by the transition away from thermally activated viscoplasticity which usually occurs at a temperature 0.8-0.9$T_{\mathrm{g}}=\gamma T_{\mathrm{g}}=T_{\mathrm{c}}(\sigma_{\mathrm{f}})=T_{\mathrm{f}}(\sigma_{\mathrm{f}})$, and a plastic transition stress of $\sigma_{\mathrm{f}}$. Finally, the sharpness of the plastic transition (set, in part, by $N$) is also an accessible experimental quantity via the kinetic fragility~\cite{Martinez2001}, which is defined as
\begin{equation}
m_{\mathrm{Kinetic}}=\left.\frac{d\log\eta}{d\left(T_{\mathrm{g}}/T\right)}\right|_{T=T_{\mathrm{g}}}. \label{EqnFragility}
\end{equation}
Here the viscosity, $\eta$ may be written via the Maxwell relation as $\eta=G_{\infty}\tau_{\mathrm{p}}(T)$, where $G$ is the instantaneous shear modulus and $\tau_{\mathrm{p}}(T)$ is now viewed as the relaxation time scale of the under-cooled liquid. The kinetic fragility gives a dimensionless measure of how rapidly the viscosity drops upon a small increase of temperature from the glass transition temperature. At this temperature one can equally well speak of a viscosity whose associated time scale is the inverse of the plastic rate. Doing so results in a simple formula for the kinetic fragility
\begin{equation}
m_{\mathrm{kinetic}}=\frac{N\overline{E}_{\mathrm{app}}(T_{\mathrm{g}})}{k_{\mathrm{b}}T_{\mathrm{g}}}, \label{EqnFragility1}
\end{equation}
where again $T_{\mathrm{c}}\approx T_{\mathrm{g}}$ is assumed. Hence the kinetic fragility sets the apparent barrier energy of the heterogeneous volume at the glass transition temperature. Eqn.~\ref{EqnFragility1} is a well known result, when viewing fragility as a kinetic phenomenon, however that it emerges from a time scale derived from the product of two exponentials whose arguments both depend on temperature (eqn.~\ref{EqnAveYieldRate4}) is not obvious and constitutes a clear justification of interpreting $\overline{E}_{\mathrm{app}}(T)$ as the apparent barrier (or activation) energy.

All of the above allow $\overline{E}$, $\delta\overline{E}(0)$ and $N$ to be determined as a function of $\overline{\alpha}$, $E_{\mathrm{p00}}$ and $\ln\left(\tau_{\mathrm{exp}}/\tau_{\mathrm{p00}}\right)$. Indeed by exploiting $T_{\mathrm{c}}\approx T_{\mathrm{g}}$, and eqns.~\ref{EqnMaxStressBroadening} and \ref{EqnFragility1}, the following relations may be obtained
\begin{eqnarray}
N\overline{E}&=&\frac{\gamma}{1-\gamma}\left[m\log10-\ln\left(\frac{\tau_{\mathrm{Exp}}}{\tau_{\mathrm{p00}}}\right)\right] k_{\mathrm{B}}T_{\mathrm{g}}-E_{\mathrm{p00}} \nonumber \\ \label{EqnCond1} \\
\sqrt{N}\delta\overline{E}(0)&=&\left[\frac{2\gamma-1}{1-\gamma}m\log10-\frac{\gamma}{1-\gamma}\ln\left(\frac{\tau_{\mathrm{Exp}}}{\tau_{\mathrm{p00}}}\right)\right]^{\frac{1}{2}}k_{\mathrm{B}}T_{\mathrm{g}} \nonumber \\ \label{EqnCond2} \\
N\overline{\alpha}&=&\frac{1}{2\left(1-\gamma\right)}\left[m\log10-\left(2-\gamma\right)\ln\left(\frac{\tau_{\mathrm{Exp}}}{\tau_{\mathrm{p00}}}\right)\right]. \nonumber \\ \label{EqnCond3}
\end{eqnarray}
$\sigma_{0}$ is given by
\begin{equation}
\frac{\sigma_{\mathrm{f}}}{\sigma_{0}}=\left[\sqrt{2N\overline{\alpha}}\gamma\frac{k_{\mathrm{B}}T_{\mathrm{g}}}{\sqrt{N}\delta\overline{E}(0)}-1\right]^{\frac{1}{2}}, \label{EqnCond4}
\end{equation}
where typically, $\sigma_{\mathrm{f}}$, is equal to $0.02G$ with $G$ being a representative (pure shear) elastic modulus. 

\begin{figure}
\begin{center}
\includegraphics[clip,width=0.9\textwidth]{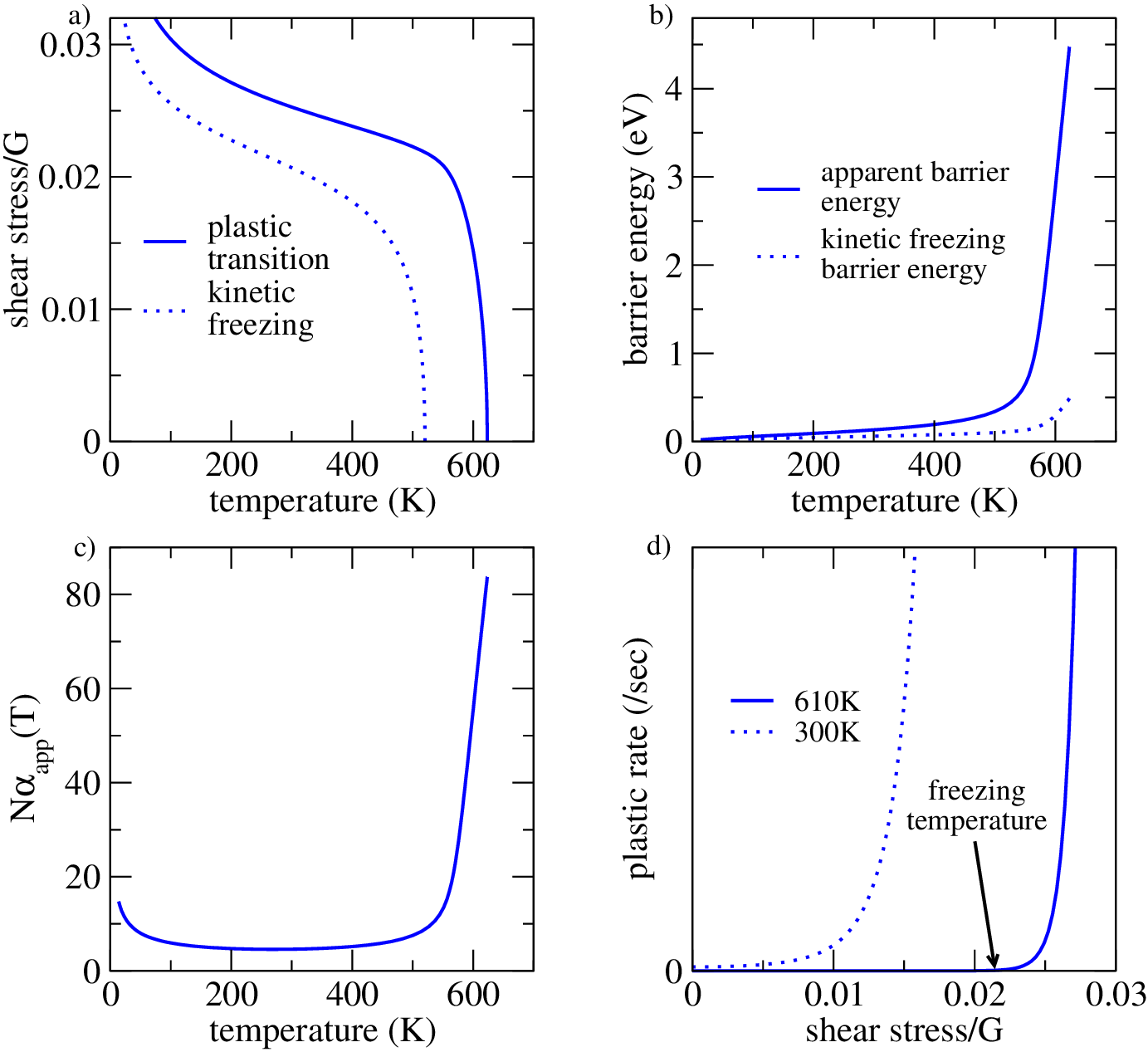}
\end{center}
\caption{a) Plastic transition and kinetic freezing stresses as a function of temperature, b) the apparent and kinetic freezing barrier energy per homogeneous volume as a function of temperature, c) Apparent number of available structural transformations as a function of temperature and d) the plastic transition rate as a function of applied shear stress for the temperatures 300K and 610K.} \label{FigExpCurves} 
\end{figure}

The above approach is motivated by the known strong correlation between the characteristic barrier energy scale of the $\alpha$ relaxation mode (derived by applying an Arrhenius viscosity law to eqn.~\ref{EqnFragility} and using experimental values for $T_{\mathrm{g}}$ and $m_{\mathrm{kinetic}}$) and the activation energy of a simple thermal activation model of plasticity that is able to describe well the high temperature ($T>0.8T_{\mathrm{g}}$) deformation properties of glasses~\cite{Schuh2007,Spaepen1977,Argon1979,Wang2011}. The above is also consistent with the initial assertion that the mega-basin barrier energy landscape is associated with the $\alpha$ relaxation mode. Furthermore, by associating $E_{\mathrm{p00}}$ with the $\beta$ relaxation mode, this material parameter may also be directly determined from experimental measurements of the $\beta$ barrier energy via differential-scanning-calorimetry or dynamical-mechanical-spectroscopy methods~\cite{Pelletier2002,Wen2004,Zhao2007,Hu2009}.

The remaining parameter is the ratio $\tau_{\mathrm{exp}}/\tau_{\mathrm{p00}}$. $\tau_{\mathrm{exp}}$ is related to a (sufficiently low) reference strain-rate, $\dot{\varepsilon}$. To first order, their relationship will be $\tau_{\mathrm{Exp}}\propto\left[\dot{\varepsilon}\right]^{-1}$, where the proportionality constant will depend on the details of the characteristic strain associated with a meta-basin escape and on temperature. These aspects, {\em  vis \'{a} vis} a stress-strain relation, will not be considered in detail in the present work, however appendix~\ref{AppDef} gives a simple estimate of the proportionality constant resulting in the simplified relation $\tau_{\mathrm{Exp}}\simeq0.01\left[\dot{\varepsilon}\right]^{-1}$. Since, the escape from a mega-basin is assumed to be mediated by $\beta$ mode relaxation, the $\tau_{\mathrm{p00}}$ will be some multiple of the system's Debye frequency. Presently it will be taken as $\tau_{\mathrm{p00}}=1\times10^{-11}$/sec.  Fortunately, it is the logarithm of the ratio of $\tau_{\mathrm{exp}}/\tau_{\mathrm{p00}}$ that enters into the model, making the model quite insensitive to the precise orders of magnitude of these timescales.

\begin{table}[ht]
\begin{center}
\begin{tabular}{cc}
\hline
Parameter & Taken experimental value \\
\hline
$T_{\mathrm{g}}$   & 623 K \\
$m_{\mathrm{kinetic}}$ &  50 \\
$E^{\alpha}(=\ln(10)m_{\mathrm{kinetic}}T_{\mathrm{g}})$ & 6.1 eV \\
$E^{\beta}$ & 1.43 eV\\
Shear modulus   &  34.1 GPa \\
\hline
\end{tabular}
\caption{Parameters taken from refs.~\cite{Lu2003,Wang2011} of Vitreloy-1 used to determine the model parameters.}
\label{Tab1}
\end{center}
\end{table}

Fig.~\ref{FigExpCurves}a shows the yield stress versus temperature for the experimental quantities of Vitreloy-1 (see tab.~\ref{Tab1}) at a strain rate of $\dot{\varepsilon}=10^{-5}$/sec, which is assumed to give a zero-load transition temperature close to the glass transition temperature. In these curves, $\gamma=0.9$ and $Na=0.125$ eV. Inspection of this figure shows that the plastic transition stress rises rapidly with decreasing temperature in the regime $T>0.9T_{\mathrm{g}}$, whereas for $T<0.9T_{\mathrm{g}}$ a change in behaviour occurs resulting in a weaker, approximately linear increase in yield stress with decreasing temperature. These curves show that at the plastic transition stress, the phenomenon of freezing is clearly avoided. Fig.~\ref{FigExpCurves}b displays the apparent barrier energy as a function of temperature. At the zero-load limit, the apparent barrier energy has a value that is equal to the experimental value of $E^{\alpha}-E^{\beta}=m_{\mathrm{kinetic}}k_{\mathrm{B}}T_{\mathrm{g}}\ln\left(10\right)-E_{\mathrm{p00}}=6.1-1.43$ eV. With decreasing temperature/increasing yield stress, the apparent barrier energy rapidly reduces until the low temperature regime is reached. In this regime, the plastic transition stress increases approximately linearly with respect to decreasing temperature.  Fig.~\ref{FigExpCurves}b also displays the freezing barrier energy. Both barrier energies remain positive for all temperatures. Fig.~\ref{FigExpCurves}c shows the natural logarithm of the apparent number of available structural transformations, $\overline{\alpha}_{\mathrm{app}}(T)N$ and, as with the apparent barrier energy, its value rapidly reduces and saturates as the low temperature regime is entered. In this regime, its value changes little. 

In the low temperature regime of deformation, experiment also reveals a robust elastic regime of stresses and a sharp transition to plasticity --- behaviour which is also seen in the present model. Fig.~\ref{FigExpCurves}d plots the plastic rate as a function of applied shear stress for the two temperatures 300K and 610K, where the latter is clearly in the high temperature regime of deformation where no kinetic freezing transition stress regime is encountered, and there is a more gradual transition to plasticity. However at a temperature of 300K, much of the stress regime below the plastic transition stress occurs within the kinetic freezing regime and the transition to plasticity is considerably sharper.

All curves are found to be independent of $\overline{\alpha}$. What is the origin of this independence? Inspection of eqns.~\ref{EqnCond1} to \ref{EqnCond3}, reveals the quantities on the left hand side are fully defined via the experimental parameters of the right hand side. Thus the mean $N\overline{E}$ (eqn.~\ref{EqnCond1}) and standard deviation $\sqrt{N}\delta\overline{E}$ (eqn.~\ref{EqnCond2}), and thus the Gaussian part of the distribution, are fixed by measurable deformation properties and independent of $\overline{\alpha}$. On the other hand, choosing an actual value of $\overline{\alpha}$ will fix $N$ via eqn.~\ref{EqnCond3} and therefore determine $\overline{E}$ and $\delta\overline{E}$.

\begin{figure}
\begin{center}
\includegraphics[clip,width=0.9\textwidth]{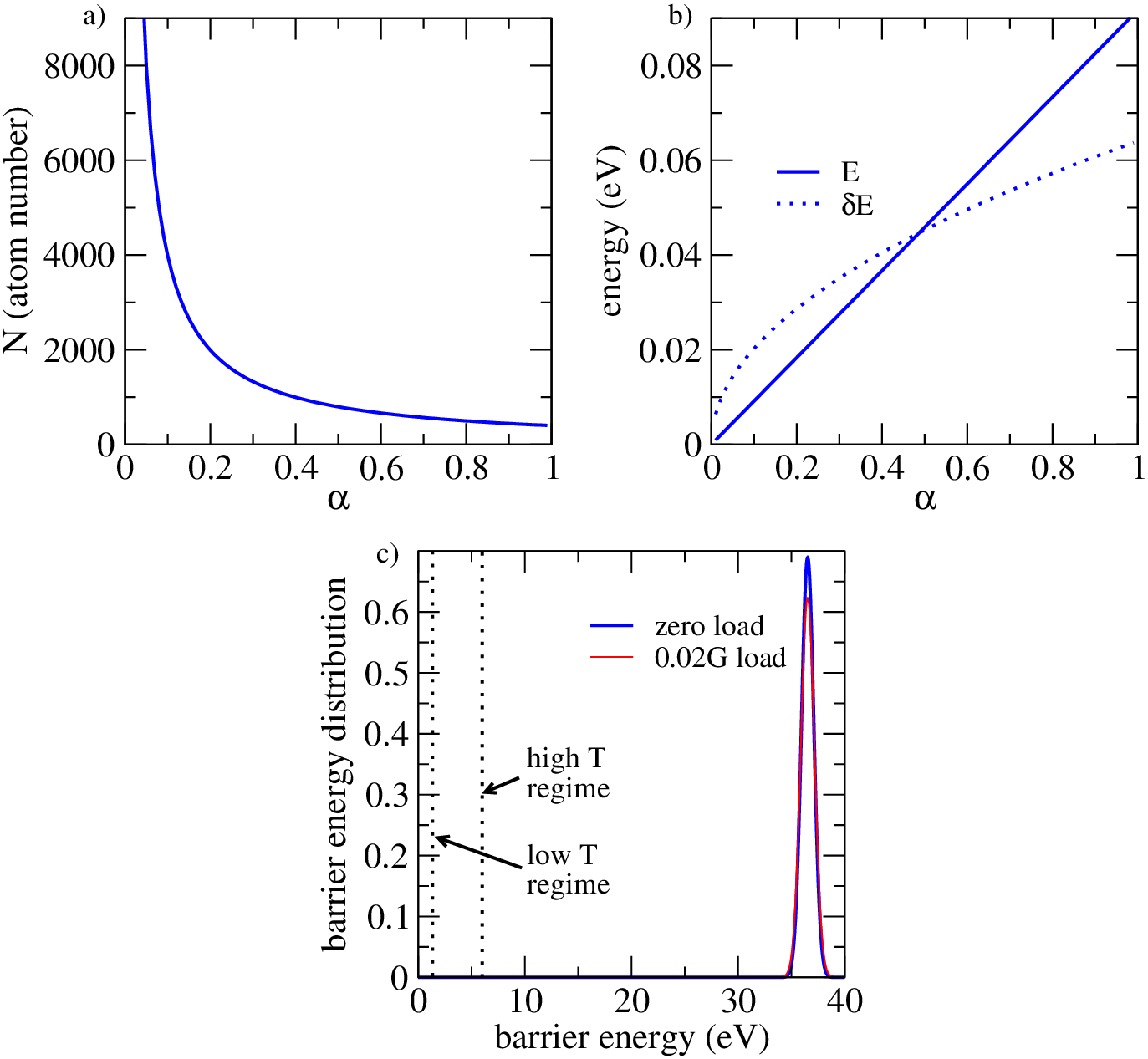}
\end{center}
\caption{a) Plot of $N$, the characteristic number of atoms per heterogeneous volume and b) the mean and standard deviation per atom of the barrier distribution. Both figures are derived from the parameters of Vitreloy-1, tab.~\ref{Tab1}. c) Barrier energy distribution for the unloaded state and at a applied stress of $0.02G$, for the parameter set corresponding to Viterol-1.} \label{FigExpParam} 
\end{figure}

Thus the experimental temperature dependence of the plastic transition stress uniquely determines the underlying distribution of $\alpha$-relaxation barrier energies, whereas $\overline{\alpha}$ sets the value of $N$, and therefore the size of the heterogeneous volume. This must be viewed as $\overline{\alpha}$ setting $N$, such that $M=\exp(\overline{\alpha}N)$ is of a sufficient size for the observed phenomenon to occur. It is in this way that the deformation properties are insensitive to the underlying microscopic detail defined via $\overline{\alpha}$. Fig.~\ref{FigExpParam}a plots eqn.~\ref{EqnCond3} and shows that for values of $\overline{\alpha}<1$, $N$ can be of the order of several thousand atoms, giving corresponding values for $\overline{E}$ and $\delta\overline{E}$ in the tens of meV per atom (fig.~\ref{FigExpParam}b). This gives a mean barrier energy of $N\overline{E}=36.5$ eV and standard deviation $\sqrt{N}\delta\overline{E}=1.3$ eV. Fig.~\ref{FigExpParam}c plots the resulting $\alpha$-relaxation barrier energy distribution, also showing the regime of barriers accessed for $T<T_{\mathrm{g}}$, as seen in fig.~\ref{FigExpCurves}b. Inspection of this figure demonstrates that the corresponding deformation properties probe only the extreme low barrier energy tail of the distribution, with the most probable part of the distribution (giving the dominant contribution to $M$) playing little role. Thus in the low temperature regime, the thermally accessible $\alpha$-relaxation PEL has significantly flattened, with the coarse grained barrier energy scale being comparable to that of the characteristic $\beta$-relaxation/STZ barrier energy.

\begin{figure}
\begin{center}
\includegraphics[clip,width=0.6\textwidth]{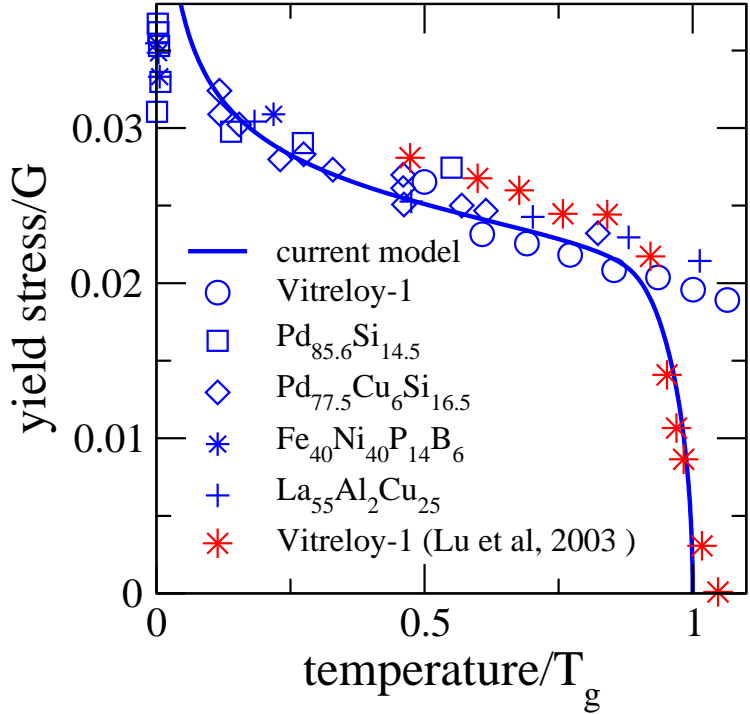}
\end{center}
\caption{Replot the plastic transition stress as a function of temperature in fig.~\ref{FigExpCurves}a, where temperature is scaled with respect to the glass transition temperature. Experimental data for Viteroly-1 and other bulk metalic glasses are also shown (blue data taken from Johnson and Samwer~\cite{Johnson2005} and red data from Lu {\em et al} \cite{Lu2003}).} \label{FigExpData} 
\end{figure}

Fig.~\ref{FigExpData} displays the typical experimental yield stress versus temperature data used to determine the optimal value of $a$. In this figure, the data from both experiment and the present theory is plotted as a yield stress (scaled by a representative shear modulus) versus temperature (scaled by the glass transition temperature). Shown is experimental data for Vitreloy-1 and other bulk metallic glasses, taken from Johnson and Samwer~\cite{Johnson2005} (presumably) at a number of different strain rates. Also displayed is Vitreloy-1 data at a strain rate of $1\times10^{-4}$/sec spanning both the low and high temperature regimes of deformation~\cite{Lu2003}. The figure shows good agreement between the current model and that of the Vitreloy-1 experimental data, and also emphasises the experimentally observed universal temperature dependence of the low temperature deformation regime. Due to the overly simple model for plastic strain (appendix~\ref{AppDef}) used to obtain a relation between the characteristic experimental time scale and the strain rate, the strain rate dependence of the developed model will not be considered in the present work.

\section{Discussion} \label{SecDis}

The present model has some similarities with the well known thermodynamic model of Adam and Gibbs \cite{Adam1965} for undercooled liquids. Both assume an exponential number of available structural transformations, although in the case of Adam and Gibbs, it is assumed that this number is equal to the number of accessible final states of the system, allowing $s_{\mathrm{config}}(T)$ in $\exp\left(s_{\mathrm{config}}(T)N^{*}\right)$ to be associated with the configurational entropy per particle. Here $N^{*}$ is the size of the cooperatively rearranging regions (CCR) whose temperature dependent value is set via the assumption that $s_{\mathrm{config}}(T)N^{*}$ is equal to a given fixed value. By assuming that the activation energy is proportional to the size of the CCR, and therefore $N^{*}$, the activation energy becomes inversely proportional to the configurational entropy per particle leading upon further assumptions to the Vogel-Tammann-Fulcher equation~\cite{Vogel1921,Tammann1926,Fulcher1925} for the characteristic relaxation time of an undercooled liquid.

For the presently developed model, the analogous $\overline{\alpha}$ is a fixed material constant and is referred to as the configurational barrier entropy per atom, $N$ is the characteristic number of atoms within a heterogeneous volume element, and the product $\overline{\alpha}N$ is set by the overall experimental temperature dependence of the plastic transition (yield) stress close to the glass transition (eqn.~\ref{EqnCond3}) as are the first and second cumulants of the barrier energy distribution (eqns.~\ref{EqnCond1} and \ref{EqnCond2}). Here, the apparent barrier energy is derived from the thermally accessible part of the barrier energy distribution, and although this barrier energy is also dependent on $N$, its temperature dependence is quite different from Adam and Gibbs, decreasing rather than increasing as the temperature drops. Indeed, in the present work, the rapid increase in the relaxation timescale (the inverse of the plastic transition rate) as the temperature decreases arises from the strong reduction in the apparent number of states and therefore $\overline{\alpha}_{\mathrm{app}}$. From this perspective a more useful comparison might be between $s_{\mathrm{config}}(T)$ and $\overline{\alpha}_{\mathrm{app}}(T)$, since both decrease with temperature and both are ultimately responsible for the rapid increase in the relaxation timescale as the temperature reduces. 

The present model may be partly viewed as a derivation of a temperature dependent critical barrier energy given by $E_{\mathrm{p00}}+N\overline{E}_{\mathrm{app}}(\sigma,T_{\mathrm{f}})$ which has contributions arising from both the $\alpha$ and $\beta$ (STZ) relaxation modes. Such a critical barrier energy has been assumed by both Argon~\cite{Argon1979} and Johnson and Samwer~\cite{Johnson2005}, although with no explicit temperature dependence, allowing for the use of a single barrier energy. Indeed for the present model, the derived temperature dependence is weak when compared to that of the high temperature deformation regime (figs.~\ref{FigYieldTempMod}c and \ref{FigExpCurves}c). Together eqns.~\ref{EqnBEMoment1} and \ref{EqnBEMoment2} (the first and second moments of the distribution $P(E)\exp(-\beta E)$) show that the $\alpha$-relaxation mode contribution to such a critical barrier energy has standard deviation that scales as $\sqrt{N}$, where $N$ is the characteristic number of atoms associated with the heterogeneous volume-scale of a BMG. For small enough $\overline{\alpha}$, $N$ can be of the order of a few thousand, resulting in $N\overline{E}_{\mathrm{app}}(\sigma,T_{\mathrm{f}})$ being a statistically meaningful quantity and therefore justifying the notion of a single critical barrier energy. Although eqns.~\ref{EqnBEMoment1} and \ref{EqnBEMoment2} are formally only applicable to the high temperature regime where a simple Gaussian suffices, numerically it was found that this is also the case for the low temperature regime of deformation when using the modified Gaussian distribution.

The current work, shows that in addition to a critical (apparent) barrier energy, there also exists a critical (apparent) number of available structural transformation and together these probe the extreme low energy tail of the barrier energy distribution (fig.~\ref{FigExpParam}c). When $T>T_{\mathrm{f}}$, this extreme low energy tail is characterised directly by the thermal distribution, $P(E)\exp(-\beta E)$, however for temperatures at and below the kinetic freezing temperature, $T_{\mathrm{f}}$, the statistics changes. In this regime, one barrier energy scale dominates and is equal to $N\overline{E}_{\mathrm{app}}(T_{\mathrm{f}})$. This kinetic freezing barrier energy is an average value, and when sampling heterogeneous volumes a distribution of such kinetic freezing barrier energies will be obtained. For positive valued distributions such as $P(E)$ the Fisher-Tippett-Gnedenko (FTG) theorem~\cite{Fisher1928,Gnedenko1943} states that such an extreme value distribution will be of the Weibull form~\cite{Gumbel2004}. Such a theorem for extremal values is analogous to the well-known central limit theorem for averages, and does not depend on the precise form of $P(E)$. The connection between the phenomenon of freezing and extreme value statistics has been established by Bouchard and Mez\'{a}rd~\cite{Bouchaud1997}, demonstrating that like the statistics of the extreme, the statistics of freezing belongs to a particular universality class suggesting a robustness against material specifics. Appendix \ref{AppEVS} details a similar connection between the present kinetic freezing phenomenon and extreme value statistics, showing that the freezing barrier energy derived in sec.~\ref{SecMD} is actually an underestimate of the extreme value statistics derived freezing barrier energy --- their difference arising from the finite size of the heterogeneous volume.

The two distinct temperature regimes of deformation exhibit clear differences in temperature dependence. What is the nature of their difference in terms of the underlying plastic activity? In the high temperature regime, each heterogeneous volume will contain a sufficient number of thermally accessible structural transformations such that plastic activity always occurs --- the degree of plasticity over the experimental time-scale $\tau_{\mathrm{exp}}$ of any sampled heterogeneous volume is well reflected by the derived average. In this regime, all heterogeneous volumes are deforming resulting in the homogeneous onset of plasticity. 

To gain insight into how the low temperature regime differs from the above scenario, the kinetic freezing limit is first considered. At a low temperature, $T$, and at an applied stress, $\sigma$, close to the freezing stress, the average plastic rate is given by 
\begin{equation}
\tau_{\mathrm{p}}\simeq\tau_{\mathrm{p00}}\times\exp\left(-\frac{E_{\mathrm{p00}}+N\overline{E}_{\mathrm{app}}(\sigma,T_{\mathrm{f}})}{k_{\mathrm{B}}T}\right),
\end{equation}
which represents the effect of only {\em one} available structural transformation within the heterogeneous volume. In this regime, sampling a particular heterogeneous volume over the time period $\tau_{\mathrm{exp}}$ can result (with a particular probability) in that volume not deforming. Thus the (negligible) plasticity at the kinetic freezing limit becomes strongly heterogeneous when sampling heterogeneous volumes. Raising the applied stress exits the kinetic freezing regime and the plastic rate is now given as
\begin{equation}
\tau_{\mathrm{p}}\simeq\tau_{\mathrm{p00}}\times M_{\mathrm{app}}(\sigma,T)\times\exp\left(-\frac{E_{\mathrm{p00}}+N\overline{E}_{\mathrm{app}}(\sigma,T)}{k_{\mathrm{B}}T}\right).
\end{equation}

Inspection of figs.~\ref{FigExpCurves}b and c reveal that at room temperature for Vitreloy-1, $M_{\mathrm{app}}(\sigma,T)$ is of the order of 100, whereas $\overline{E}_{\mathrm{app}}(\sigma,T)$ has not increased greatly from that of its freezing value. Although plasticity is more likely in any sampled heterogeneous volume at the actual plastic transition stress, in transiting to non-negligible plasticity, the statistics is expected to be strongly influenced by those of the kinetic freezing regime. 

Thus the transition from the high temperature regime to the low temperature regime is characterised by a change of statistics from that where the average plastic time-scale reflects well the degree of plasticity of any sampled heterogeneous volume to that where the statistics of small numbers and extremal values admit a non-negligible probability of any particular heterogeneous volume not deforming. The implication is that in passing from the high to low temperature regime, the plasticity becomes inherently inhomogeneous in the transition from elasticity to plasticity --- the so-called micro-plastic regime of deformation. How such a low temperature heterogeneity manifests itself as an emerging material instability in the macro-plastic/flow via (say) shear banding would give fundamental insight into the transition from homogeneous to heterogeneous plasticity seen experimentally as the temperature decreases from the glass transition.

The generality of the current result is now discussed. Whilst the use of a modified Gaussian is certainly an approximation to the true distribution of $\alpha$-relaxation barrier energies, the existence of both a kinetic freezing and plastic transition temperature is insensitive to the precise form of $P(E)$, where its essential properties are the extensive first and second cumulants --- a requirement that is intimately related to the known exponential scaling of the number of stationary configurations of a structural glass. Moreover, at and close to the freezing regime, where extreme value statistics dominate, the resulting Weibull distribution via the FTG theorem is a universal result independent of the precise form of $P(E)$. Thus, while other barrier energy distributions which have extensive first and second order cumulants can certainly be considered such as a log-normal distribution which has the correct limit $P(E\rightarrow0)\rightarrow0$, the general result of the present work is expected to be insensitive to the specific choice of $P(E)$.

One implicit approximation thus far not discussed, is that the average over heterogeneous volumes is taken with the assumption that each such volume is statistically independent of the other, that is, they do not interact. A similar assumption is made for the CCR of the Adam and Gibbs formalism~\cite{Adam1965}. The assumption of a lack of an explicit interaction between heterogeneous volumes is clearly an approximation to make the problem analytically simple. The universal phenomenon of freezing has however been found to be robust against certain forms of interactions~\cite{Carpentier2001,Fyodorov2008}, and it is expected that for the low temperature regime where kinetic freezing plays a defining role in determining the nature of plasticity, the current results will not be fundamentally changed by the inclusion of interactions between the heterogeneous volumes. It remains a topic of future work to, in the first instance, characterise the interaction between heterogeneous volume elements and then to implement numerical procedures that are able to investigate their role not only in the transition from elasticity to plasticity but also in the flow regime of macro-plasticity.

The Arrhenius form used for $\tau_{\mathrm{p0}}$, eqn.~\ref{EqnYieldRate3}, to describe the average properties of the atomic scale $\beta$-relaxation/STZ processes mediating the $\alpha$-relaxation landscape exploration is the simplest choice that embodies the thermal activation hypothesis. More complex temperature dependencies could be envisaged emerging from the distribution of barrier energies known to also exist for $\beta$-relaxation barrier energies derived from atomic scale PEL explorations~\cite{Rodney2009a,Rodney2009b,Koziatek2013,Kallel2010} using the ART$n$ technique~\cite{Barkema1996,Mousseau1998,Olsen2004}. Due to the independence of such phenomena to the size of the heterogeneous volume, $N$, sec.~\ref{SecAER} shows however that any such temperature dependence will only renormalize the parameters of the current model to a leading order of $1/N$. Thus the present model should be insensitive to the precise form of $\tau_{\mathrm{p0}}$ whether it be a function of temperature and/or applied stress.

The current work has only considered an applied stress that is a simple shear. In comparing to experiment, it was found that the a representative shear stress modulus sets the proportionality constant between the applied shear stress and the broadening of $P(E)$, eqn.~\ref{EqnCond4}. Deformation experiments are however usually performed under uni-axial loading conditions, were a compressive/dilatory component is also present. A more complex applied stress that contains such an isotropic component, such as the uni-axial loading experiment is expected to affect $P(E)$ in a more complex manner by depending not only on a pure shear modulus but also on the Poisson ratio. This aspect, and thus the role of local volume changes, will be investigated in future work.

\section{Concluding Remarks}

In the 2000 review article of Angell {\em et al} \cite{Angell2000}, which articulated the contemporary questions to be addressed in the field of structural glasses, one such question for the low temperature regime was ``What types of processes remain active in the glass when the $\alpha$ relaxation has been completely frozen?''. The present work demonstrates how such a freezing of the $\alpha$ relaxation potential energy landscape could occur as the temperature reduces. Indeed as the temperature decreases an increasing portion of this landscape becomes thermally inaccessible and therefore frozen --- a process that is only complete at a temperature of absolute zero. Thus the statistics of the $\alpha$ relaxation processes, so integral to the under-cooled liquid regime, are also found to play a central role in determining those thermally active processes that lead to low temperature plasticity.

In summary, a model has been developed to describe the temperature dependence of the transition from elasticity to plasticity, the micro-plastic regime, in a bulk metallic glass. Central to this model, is the assumption of thermally activated plasticity, and in particular, that plasticity occurs via the thermal activation of $\alpha$-relaxation processes that are themselves mediated by multiple thermally activated $\beta$-relaxation/STZ activity. Further, it is assumed that the known exponentially number of stationary points in a glassy potential energy landscape for the under-cooled liquid regime, results in a distribution of $\alpha$-relaxation barrier energies that has first and second cumulants which are extensive with respect to the underlying heterogeneous volume-scale of the glass. 

Two distinct temperature regimes emerge. At high temperature, the shear stress at which significant plasticity occurs rises rapidly from its zero value close to the glass transition temperature, corresponding to a drop in the apparent barrier energy and apparent number of available structural transitions. In the case of the latter, this reduction is comparable to the many orders of magnitude seen in the increase in the viscosity as the glass transition temperature is approached from the under-cooled liquid regime. At a low enough temperature, the apparent number of structural transformations saturates, due to the eventual onset of kinetic freezing, and the low temperature regime is entered where the plastic transition stress now increases approximately linearly with respect to decreasing temperature. In this deformation mode, the transition from elasticity to plasticity is sharp with respect to the applied stress, and for applied stresses characteristic of the elastic regime plastic activity is kinetically frozen indicating a single dominant barrier energy resulting in an underlying robustness of the elastic regime.

\section{Acknowledgements}

PMD thanks D. Rodney for valuable discussions and RM gratefully acknowledges the financial support of the Alexander von Humboldt foundation.

\appendix

\section{An analogy to thermodynamics and the phenomenon of freezing} \label{AppTherm}

Further insight into eqn.~\ref{EqnAveYieldRate3} can be obtained by recognising that 
\begin{equation}
M\left\langle\exp\left(-\frac{E}{k_{\mathrm{B}}T}\right)\right\rangle=
\left\langle\sum_{i=1}^{M}\exp\left(-\frac{E_{i}}{k_{\mathrm{B}}T}\right)\right\rangle,\label{EqnPFApp}
\end{equation}
where $\langle\cdots\rangle$ is an average over heterogeneous volumes, has the structure of an environmentally averaged partition function average. Thus the statistics of the current barrier energy problem can be mapped to an equilibrium statistical mechanics framework, which then allows for application of Derrida's micro-canonical approach to the phenomenon of freezing \cite{Derrida1980}.

The average number of barrier energies between $E$ and $E+dE$, $\langle\Omega(E)\rangle$, is given by 
\begin{equation}
\langle\Omega(E)\rangle=M(E)dE=MP(E)dE, \label{EqnNumDen}
\end{equation}
where $dE$ must be small enough to ensure a well defined barrier energy but also large enough so that $\langle\Omega(E)\rangle$ is a smooth function of $E$ (see ref. \cite{Derrida1980} for a related discussion). For a sufficiently large heterogeneous volume, fluctuations in eqn.~\ref{EqnNumDen} become small and $\langle\Omega(E)\rangle$ becomes a statistically meaningful quantity, allowing for a corresponding barrier energy to be defined via $S(E)=k_{\mathrm{B}}\ln\langle\Omega(E)\rangle$. For a Gaussian distribution with mean $N\overline{E}$ and standard deviation $\sqrt{N}\delta\overline{E}$ this is
\begin{eqnarray}
S(E)&=&k_{\mathrm{B}}N\left(\overline{\alpha}-\frac{1}{2}\left(\frac{E/N-\overline{E}}{\delta\overline{E}}\right)^{2}\right)\\ \nonumber
& &+k_{\mathrm{B}}\ln\left(\frac{dE}{\sqrt{2\pi N}\delta\overline{E}}\right).\label{EqnEntropyApp}
\end{eqnarray}
From equilibrium thermodynamics, an effective temperature is formally defined via
\begin{equation}
\left.\frac{dS(E)}{dE}\right|_{E=U(T)}=\frac{1}{T}, \label{EqnThermTempApp}
\end{equation}
and from eqn.~\ref{EqnEntropyApp}, this gives,
\begin{equation}
U(T)=N\left(\overline{E}-\frac{\left(\delta\overline{E}\right)^{2}}{k_{\mathrm{B}}T}\right)\label{EqnInternalEnergyApp}
\end{equation}
as the relation between internal barrier energy and effective temperature. It is noted that the internal barrier energy, $U(T)$ from eqn.~\ref{EqnInternalEnergyApp} is equal to the apparent barrier energy of eqn.~\ref{EqnBEMoment1}. With the barrier entropy (eqn.~\ref{EqnEntropyApp}) and the barrier internal energy (via eqn.~\ref{EqnInternalEnergyApp}), the free barrier energy, $F(T)=U(T)-S(T)T$, can be constructed:
\begin{eqnarray}
F(T)&=&U(T)-k_{\mathrm{B}}TN\left(\overline{\alpha}-\frac{1}{2}\left(\frac{U(T)/N-\overline{E}}{\delta\overline{E}}\right)^{2}\right)\nonumber\\
&=&-k_{\mathrm{B}}TN\left(\overline{\alpha}-\frac{\overline{E}}{k_{\mathrm{B}}T}+\frac{1}{2}\left(\frac{\delta\overline{E}}{k_{\mathrm{B}}T}\right)^{2}\right), \label{EqnFreeEnergyApp}
\end{eqnarray}
where in the last step, eqn.~\ref{EqnInternalEnergyApp} has been used to make the free barrier energy dependent on effective temperature. Since the average partition function is proportional to the logarithmic of the free barrier energy, eqn.~\ref{EqnPFApp} can be written as
\begin{equation}
M\left\langle\exp\left(-\frac{E}{k_{\mathrm{B}}T}\right)\right\rangle=\exp\left(-\frac{F(T)}{k_{\mathrm{B}}T}\right),
\end{equation}
which is identical to eqn.~\ref{EqnAveYieldRate3}. That eqn.~\ref{EqnAveYieldRate3} can be achieved in this manner, indicates that the definition of an effective temperature via eqn.~\ref{EqnThermTempApp} formally corresponds to the true temperature of the system. 

In eqn.~\ref{EqnFreeEnergyApp}, the second term of eqn.~\ref{EqnEntropyApp} has been dropped under the assumption that the heterogeneous volume is large enough such that it contributes neglibly to both the barrier entropy and the free barrier energy. This highlights the fact that the present derivation and that of sec.~\ref{SecMD} are formally equivalent only in the limit $N\rightarrow\infty$. Continuing in this limit, at a sufficiently low enough barrier energy $\left\langle\Omega(E_{\mathrm{f}})\right\rangle=1$ gives
\begin{equation}
S(E_{\mathrm{f}})=Nk_{\mathrm{B}}\left(\overline{\alpha}-\frac{1}{2}\left(\frac{E_{\mathrm{f}}/N-\overline{E}}{\delta\overline{E}}\right)^{2}\right)=0, \label{EqnFreezingCondApp}
\end{equation}
which has the relevant solution
\begin{equation}
E_{\mathrm{f}}=N\left(\overline{E}-\delta\overline{E}\sqrt{2\overline{\alpha}}\right).\label{EqnFreezingBEApp}
\end{equation}
or in terms of temperature ($E_{\mathrm{f}}=U(T_{\mathrm{f}})$) via eqn.~\ref{EqnThermTempApp},
\begin{equation}
T_{\mathrm{f}}=\frac{\delta\overline{E}}{k_{\mathrm{B}}}\frac{1}{\sqrt{2\overline{\alpha}}}. \label{EqnFreezingTempApp}
\end{equation}
Eqns.~\ref{EqnThermTempApp} and \ref{EqnFreezingTempApp} are identical to those derived for the freezing barrier energy and temperature in sec.~\ref{SecGauss}.

From the perspective of thermodynamics, at the temperature $T_{\mathrm{f}}$ the entropy is equal to zero and since upon further decrease in temperature the thermodynamic entropy cannot further reduce, the system freezes into this single dominant state resulting in a temperature independent free energy~\cite{Derrida1980}. For the present analogy to barrier energy kinetics, there is no {\em a priori} reason why the barrier entropy cannot become negative, however the discussion in sec.~\ref{SecGauss}, in terms of extreme value statistics, indicates the analogy to thermodynamics may be also applied to the phenomenon of freezing. This has been demonstrated by Bouchard and M\'{e}zard~ \cite{Bouchaud1997} where the freezing phenomenon could be understood from the perspective of extreme value statistics. These aspects are also briefly considered in appendix C.

Alternatively, the entire procedure entailed by eqns.~\ref{EqnEntropyApp} to \ref{EqnFreeEnergyApp} can be viewed as a change of integration variables resulting in  the integral of eqn.~\ref{EqnAveYieldRate3} in sec.~\ref{SecGauss} being transformed to a contour integral in the complex plane. This approach becomes advantageous when considering the modified Gaussian of sec.~\ref{SecGaussM}, which gives the barrier entropy as
\begin{equation}
S(E)=k_{\mathrm{B}}N\left(\overline{\alpha}-\frac{1}{2}\left(\frac{g(E)/N-\overline{E}}{\delta\overline{E}}\right)^{2}\right), \label{EqnEntropyMod}
\end{equation}
and the relation between internal barrier energy and temperature as
\begin{equation}
\left.-k_{\mathrm{B}}\left(\frac{g(E)/N-\overline{E}}{\delta\overline{E}^{2}}\right)g'(E)\right|_{E=U(T)}=\frac{1}{T}. \label{EqnInternalEnergyMod}
\end{equation}
To obtain the internal energy as a function of temperature, and thereby be able to construct the free barrier energy as a function of temperature, eqn.~\ref{EqnInternalEnergyMod} must be inverted. For particular forms of $g(E)$ this may be done analytically, but more generally, such an inversion is  done numerically. For the case of the barrier energy at freezing, $S(E_{\mathrm{f}})=0$ leads to
\begin{equation}
E_{\mathrm{f}}=Ng^{-1}\left(\overline{E}-\delta\overline{E}\sqrt{2\overline{\alpha}}\right).
=Ng^{-1}\left(\overline{E}_{\mathrm{f}}^{\mathrm{G}}\right) \label{EqnFreezingBarrierApp}
\end{equation}
where $\overline{E}_{\mathrm{f}}^{\mathrm{G}}$ is the freezing barrier energy derived from a pure Gaussian distribution (eqn.~\ref{EqnAppEF}).

\section{A simple deformation model} \label{AppDef}

To gain an estimate of the relationship between $\tau_{\mathrm{exp}}$ and the strain rate  $\dot{\varepsilon}_{\mathrm{total}}$ the following simplified viscoplastic model is developed. Assuming an additive elastic and plastic strain rate, the total strain rate is given by
\begin{equation}
\dot{\varepsilon}_{\mathrm{total}}=\dot{\varepsilon}_{\mathrm{e}}+\dot{\varepsilon}_{\mathrm{p}}(\sigma)=\frac{\dot{\sigma}}{G}+\dot{\varepsilon}_{\mathrm{p}}(\sigma),
\end{equation}
which gives the first order equation
\begin{equation}
\frac{d\sigma}{dt}=G\left[\dot{\varepsilon}_{\mathrm{total}}-\dot{\varepsilon}_{\mathrm{p}}(\sigma)\right],
\end{equation}
whose solution may be written as an integral equation
\begin{equation}
\sigma(t+\delta t)=\sigma(t)+G\left[\dot{\varepsilon}_{\mathrm{total}}\delta t-\int_{t}^{t+\delta t}dt\,\dot{\varepsilon}_{\mathrm{p}}(\sigma(t))\right].
\end{equation}
For a small enough time interval this may be approximated as
\begin{equation}
\sigma(t+\delta t)=\sigma(t)+G\dot{\varepsilon}_{\mathrm{total}}\delta t\left[1-\frac{\dot{\varepsilon}_{\mathrm{p}}(\sigma(t))}{\dot{\varepsilon}_{\mathrm{total}}}\right],
\end{equation}
which can be iterated to give a, constant strain-rate, stress-strain curve. Inspection of the above equation gives the condition for perfect plastic flow ($\sigma(t+\delta t)=\sigma(t)$) as when the plastic strain rate is equal to the total applied strain rate ($\dot{\varepsilon}_{\mathrm{p}}=\dot{\varepsilon}_{\mathrm{total}}$).

To obtain an estimate for $\dot{\varepsilon}_{\mathrm{p}}$, and therefore a connection to  $\dot{\varepsilon}_{\mathrm{total}}$, the simplest approximation is to linearly relate the plastic strain rate of a glass of volume $V$ to the presently derived plastic rate, $\left[\tau_{\mathrm{p}}\right]^{-1}$:
\begin{equation}
\dot{\varepsilon}_{\mathrm{p}}=\Delta\varepsilon\times\left[\tau_{\mathrm{p}}\right]^{-1}\times\frac{V}{N(l_{0})^{3}}.
\end{equation}
In the above, $l_{0}$ is a measure of the average atomic spacing, giving $V/N(l_{0})^{3}$ as the number of heterogeneous volumes within the material volume $V$, and $\Delta\varepsilon$ is an estimate of the characteristic plastic strain arising from an irreversible structural transformation occurring within a heterogeneous volume. Under the assumption that an irreversible structural transformation can be well represented by a localized shearing of the material, according to \cite{Eshelby1954, Eshelby1957}, $\Delta\varepsilon$ will be given by the ratio of the characteristic slipped area projected onto the gauge cross-section multiplied by the slip distance, and $V$. Assuming the characteristic slipped area has a length scale that is comparable to the heterogeneity length scale (say, half) and the corresponding slip distance will be of the order of an atomic spacing (since the mediating $\beta$-relaxation processes will involve $O(1)$ atoms~\cite{Stillinger1995}), the above equation reduces to
\begin{equation}
\dot{\varepsilon}_{\mathrm{total}}\sim\frac{1}{4N^{1/3}}\times\left[\tau_{\mathrm{p}}\right]^{-1}.
\end{equation}
For Vitreloy-1 the proportionality constant is approximately 0.01 when $N\sim3000$ (corresponding to $\overline{\alpha}\approx0.1$). It is noted that all geometric factors have been ignored and that the above should be viewed at best as an order of magnitude estimate, which is sufficient given that the resulting timescale enters enters the logarithmic factor in eqn.~\ref{EqnRenormAlpha}. In the present context, a choice of a corresponding characteristic plastic strain, $\Delta\varepsilon$, is actually a choice on the nature of the mega-basin the material finally enters upon exiting its current mega-basin --- an aspect that is not included in the current model. The current choice simply assumes that the system finds (on average) a final state that is compatible with the external loading geometry.

\section{The connection to extreme value statistics} \label{AppEVS}

\begin{figure}
\begin{center}
\includegraphics[clip,width=0.6\textwidth]{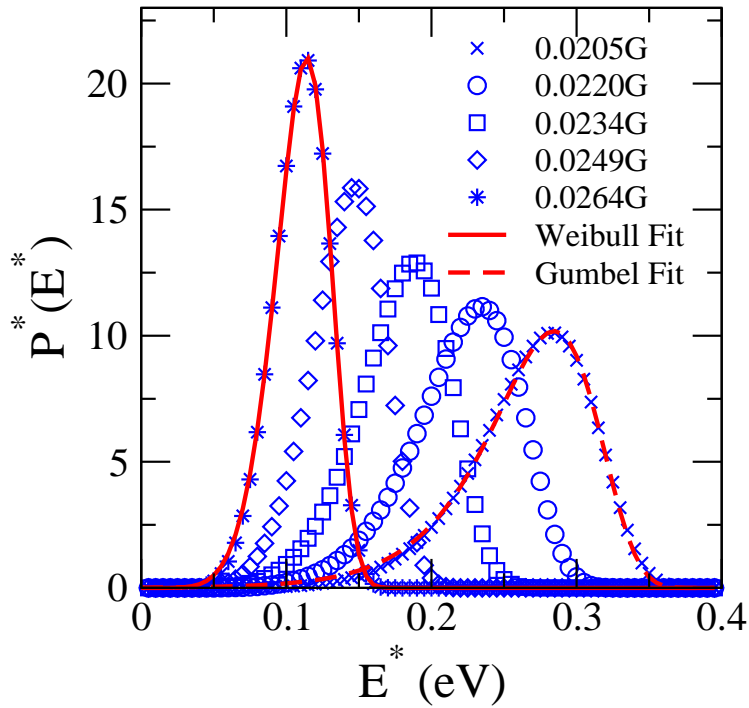}
\end{center}
\caption{Plot of extreme value barrier energy distribution derived from the barrier energy distribution for the parameter set corresponding to Vitreloy-1 (fig.~\ref{FigExpParam}c) for five stresses spanning both the low and high temperature regimes (fig.~\ref{FigExpData})). Also included are the optimal Gumbel fit for the lowest stress distribution and the optimal Weibull fit for the highest stress distribution.} \label{Fig1App3} 
\end{figure}

When sampling the barrier energy distribution $P(E)$, $M$ times, the probability that the minimum barrier energy is $E^{*}$ is given by~\cite{Bouchaud1997}
\begin{equation}
P_{\mathrm{min}}(E^{*})=M P(E^{*})\left(1-P_{<}(E^{*})\right)^{M-1}, \label{Eqn1App3}
\end{equation}
where $P_{<}(E)$ is the repartition probability or the cumulative distribution function of $P(E)$:
\begin{equation}
P_{<}(E^{*})=\int_{0}^{E^{*}}dE\,P(E). \label{Eqn2App3}
\end{equation}
Eqn.~\ref{Eqn1App3} is the required extreme value distribution associated with the statistics of the smallest barrier energy in a heterogeneous volume containing $M$ possible barrier energies. Fig.~\ref{Fig1App3} plots eqn.~\ref{Eqn1App3} using $P(E)$ derived from the parameters of Vitreloy-1 at a number of different stresses spanning both the low and high temperature plastic transition stress versus temperature curve of fig.~\ref{FigExpData}. Inspection of this figure reveals distributions that are peaked at energies close to that of the kinetic freezing barrier energy scale and with increasing stress are shifted to smaller barrier energies. The shape of the extreme value distribution also changes upon increasing the stress, where for the largest stress the distribution could be fitted optimally using a Weibull distribution, and for the lowest stress, a Gumbel distribution provided the best fit. Such a result is expected since at low stresses (the high temperature regime) the Gaussian form dominates with the extreme value statistics unaware of the necessity of a positive barrier energy, giving a Gumbel extreme value distribution. However as the stress increases (and the low temperature regime is entered), the modified Gaussian form begins to dominate resulting in a Weibull extreme value distribution.

Inspection of the corresponding mean derived from each distribution reveals values that are consistently higher than that predicted by the kinetic freezing barrier energy eqn~\ref{EqnFreezingBarrierApp}. The origin of this may be seen by determining an approximate expression for the extreme value distribution derived mean value. For sufficiently large $M$,
\begin{equation}
\left(1-P_{<}(E^{*})\right)^{M}\simeq\exp\left(-M P_{<}(E^{*})\right).\label{EqnApp}
\end{equation}
The distribution entailed by eqn.~\ref{Eqn1App3} will peak when the argument of the exponetial in the above is approximately one giving a definition for the mean minimum value, $E^{*}_{\mathrm{f}}$, as
\begin{equation}
P_{<}(E^{*}_{\mathrm{f}})=\frac{1}{M}. \label{Eqn4App3}
\end{equation}
For small enough $E$, the repartition distribution can be approximated to logarithmic accuracy as
\begin{equation}
P_{<}(E)\simeq\frac{EP(E)}{2\pi}, \label{Eqn5App3}
\end{equation}
resulting in eqn.~\ref{Eqn4App3} reducing to
\begin{equation}
\log\left(P(E^{*}_{\mathrm{f}})\right)\simeq\log\left(\frac{1}{M}\right)-\log\left(\frac{E^{*}_{\mathrm{f}}}{2\pi}\right).  \label{Eqn6App3}
\end{equation}
For the pure Gaussian form, eqn.~\ref{EqnGD}, this reduces to 
\begin{equation}
\overline{\alpha}-\frac{1}{2}\left(\frac{\overline{E}^{*}_{\mathrm{f}}-\overline{E}}{\delta\overline{E}}\right)^{2}\simeq
-\frac{1}{N}\log\left(\sqrt{\frac{N}{(2\pi)^{3}}}
\frac{\overline{E}^{*}_{\mathrm{f}}}{\delta\overline{E}}\right). \label{Eqn7App3}
\end{equation}
The above equation gives an accurate estimation of the average minimum energy barrier, $\overline{E}^{*}_{\mathrm{f}}$,  for the extreme value distributions for Vitreloy-1 shown in fig.~\ref{Fig1App3}. It is only in the bulk limit, $N\rightarrow\infty$, that eqn.~\ref{Eqn7App3} becomes identical to eqn.~\ref{EqnFreezingCondApp} resulting in $\overline{E}^{*}_{\mathrm{f}}=\overline{E}_{\mathrm{f}}$ and a formal equivalence between extreme value statistics and the thermodynamic treatment of the kinetic freezing phenomenon developed in appendix A, as was shown by Bouchaud and M\'{e}zard~\cite{Bouchaud1997}. The right-hand-side of eqn.~\ref{Eqn7App3} is therefore a correction that takes into account the finite size of the heterogeneous volume.


\begin{thebibliography}{99}
\bibitem{Schuh2007} C. A. Schuh, T. C. Hufnagel, U. Ramamurty, Acta. Mater. 55 (2007) 4067.
\bibitem{Wu2008} W.F. Wu, Y. Li and C.A. Schuh, Phil. Mag. 88 (2008), p.71.
\bibitem{Spaepen1977} F. Spaepen, Acta Metall. 25 (1977) p.407.
\bibitem{Argon1979} A. Argon, Acta Metall. 27 (1979) p.47.
\bibitem{Falk1998} M.L. Falk and J.S. Langer, Phys. Rev. E 57 (1998) p.7192.
\bibitem{Schuh2003} C.A. Schuh and A.C. Lund, Nat. Mater. 2 p.449.
\bibitem{Maloney2004} C.E. Maloney and A. Lema\^{i}tre, Phys. Rev. Lett. 93 (2004) p.016001.
\bibitem{Demkowicz2005} M.J. Demkowicz and A.S. Argon, Phys. Rev. B 72 (2005) p.245205.
\bibitem{Shi2006} Y. Shi and M.L. Falk, Phys. Rev. B 73 (2006) p.214201.
\bibitem{Rodney2009a} D. Rodney and C.A. Schuh, Phys. Rev. Lett. 102 (2009) p.235503.
\bibitem{Rodney2009b} D. Rodney and C.A. Schuh, Phys. Rev. B. 80 (2009) p.184203.
\bibitem{Rodney2011} D. Rodney, A. Tanguy and D. Vandembroucq, Modelling Simul. Mater. Sci. Eng. 19 (2011) p.083001.
\bibitem{Guan2010} P. Guan, M. Chen and T. Egami, Phys. Rev. Lett. 104 (2010) p.205701.
\bibitem{Liu1998} A. Liu and S.R. Nagel, Nature 396 (1998) p.21.
\bibitem{Trappe2001} V. Trappe, V. Prasad, L. Cipelletti, P.N. Segre and D.A. Weitz, Nature 411 (2001) p.772.
\bibitem{Bulatov1994a} V.V. Bulatov and A.S. Argon, Model. Simul. Mater. Sci. Eng. 2 (1994) p.167.
\bibitem{Bulatov1994b} V.V. Bulatov and A.S. Argon, Model. Simul. Mater. Sci. Eng. 2 (1994) p.185.
\bibitem{Bulatov1994c} V.V. Bulatov and A.S. Argon, Model. Simul. Mater. Sci. Eng. 2 (1994) p.203.
\bibitem{Homer2009} E.R. Homer and C.A. Schuh, Acta Mater. 57 (2009) p.2823.
\bibitem{Homer2010} E.R. Homer and C.A. Schuh, Modelling Simul. Mater. Sci. Eng. 18 (2010) p.065009.
\bibitem{Heggen2005} M. Heggen, F.X. Spaepen and M. Feuerbacher, J. Appl. Phys. 97 (2005) p.033506.
\bibitem{Wang2011} W.H. Wang, J. Appl. Phys. 110 (2011) p.053521.
\bibitem{Goldstein1969} M. Goldstein, J. Chem. Phys. 51 (1969) p.3728.
\bibitem{Johari1970a} G.P. Johari and M. Goldstein, J. Phys. Chem. 74 (1970) p.2034.
\bibitem{Johari1970b} G.P. Johari and M. Goldstein, J. Chem. Phys. 53 (1970) p.2372.
\bibitem{Johari1973} G.P. Johari, J. Chem. Phys. 58 (1973) p.1766.
\bibitem{Stillinger1995} F.H. Stillinger, Science 267 (1995) p.1935.
\bibitem{Debenedetti2001} P.G. Debenedetti and F.H. Stillinger, Nature 419 (2001) p.259.
\bibitem{Angell1991} C.A. Angell, J. Non-Cryst. Sol. 131-133 (1991) p.13.
\bibitem{Heuer2008} A. Heuer, J. Phys.: Condens. Matter 20 (2008) p.373101.
\bibitem{Harmon2007} J.S. Harmon, M.D. Demetriou, W.L. Johnson and K. Samwer, Phys. Rev. Lett. 99 (2007) p.135502.
\bibitem{Bouchbinder2007} D.E. Bouchbinder, J.S. Langer and I. Procaccia, Phys. Rev. E 75 (2007) p.036107.
\bibitem{Falk2011} M.L. Falk and J.S. Langer, Annual Review of Condensed Matter Physics 2 (2011) p.353.
\bibitem{Kimura1980} H. Kimura and T. Masumoto, Acta Metall. 28 (1980) p.1663.
\bibitem{Kimura1982} H. Kimura and T. Masumoto, J. Appl. Phys. 53 (1982) p.3523.
\bibitem{Kimura1983} H. Kimura and T. Masumoto, Acta Metall. 31 (1983) p.231.
\bibitem{Dubach2009} A. Dubach, F.H. Dalla Torre and J.F. L\"{o}ffler, Acta Mater. 57 (2009) p.881.
\bibitem{Klaumunzer2010} D. Klaum{\"{u}}nzer, R. Maa{\ss}, F.H. Dalla Torre and J.F. L\"{o}ffler, Appl. Phys. Lett. 96 (2010) p.061901.
\bibitem{Maass2011} R. Maa\ss, D. Klaum\"{u}zer and J.F. L\"{o}ffler, Acta Mater. 59 (2011) p.3205.
\bibitem{Maass2012} R. Maa\ss, D. Klaum\"{u}zer, G. Villard, P.M. Derlet and J.F. L\"{o}ffler, Appl. Phys. Lett. 100 (2012) p.071904.
\bibitem{Johnson2005} W.L. Johnson and K.A. Samwer, Phys. Rev. Lett. 95 (2005) p.195501.
\bibitem{Derlet2011} P.M. Derlet and R. Maa\ss, Phys. Rev. B (RC) 84 (2011) p.220201.
\bibitem{Derlet2012} P.M. Derlet and R. Maa\ss, Mater. Res. Soc. Symp. Proc. Vol. 1520, DOI: 10.1557/opl.2012.1689 (2012).
\bibitem{Argon1968} A.S. Argon, J. Appl. Phys. 39 (1968) p.4080.
\bibitem{Argon1980} A.S. Argon and H.Y. Kuo, J. Non-Crys. Sol. 37 (1980) p.241.
\bibitem{Khonik2000} V.A. Khonik, phys. stat. sol. (a) 177 (2000) p.173.
\bibitem{Gibbs1993} M.R.J. Gibbs, J.E. Evetts and J.A. Leake, J. Mat. Sci. 18 (1983) p.278.
\bibitem{Khonik1998} V.A. Khonik, A.T. Kosilov, V.A. Mikhailov and V.V. Sviridov, Acta. Mater. 46 (1998) p.3399.
\bibitem{Khonik2001} V.A. Khonik and M. Ohta, phys. stat. sol. (a) 184 (2001) p.367.
\bibitem{Primak1955} W. Primak, Phys. Rev. 100 (1955) p.1677.
\bibitem{Koziatek2013} P. Koziatek, J.-L. Barrat, P.M. Derlet, D. Rodney, Phys. Rev. B 87 (2013) p. 224105.
\bibitem{Barkema1996} G.T. Barkema and N. Mousseau, Phys. Rev. Lett. 77 (1996) p. 4358
\bibitem{Mousseau1998} N. Mousseau and G.T. Barkema, Phys. Rev. E. 57 (1998) p. 2419
\bibitem{Olsen2004} R.A. Olsen, G.J. Kroes, G. Henkelman and A. Arnaldsson, H. J\'{o}nsson, J. Chem. Phys. 121 (2004) p.9776.
\bibitem{Kallel2010} H. Kallel, N. Mousseau and F. Schiettekatte, Phys. Rev. Lett. 105 (2010) p. 045503
\bibitem{Stillinger1983} F.H. Stillinger and T.A. Weber, Phys. Rev. A 28 (1983) p.2408.
\bibitem{Stillinger1984} F.H. Stillinger and T.A. Weber, Science 225 (1984) p.983.
\bibitem{Stillinger1999} F.H. Stillinger, Phys. Rev. E 59 (1999) p.48.
\bibitem{Shell2004} M. Scott Shell, P.G. Debenedetti and A.Z. Panagiotopoulous, Phys. Rev. Lett. 92 (2004) p.035506.
\bibitem{Fyodorov2004} Y.V. Fyodorov, Phys. Rev. Lett. 92 (2004) p.240601.
\bibitem{Kohen2000} D. Kohen and F.H. Stillinger, Phys. Rev. E 61 (2000) p.1176.
\bibitem{Adam1965} G. Adam and J.H. Gibbs, J. Chem. Phys. 43 (1965) p.139.
\bibitem{Kirkpatrick1989} T.R. Kirkpatrick, D. Thirumalai and P.G. Wolynes, Phys. Rev. A 40 (1989) p.1045.
\bibitem{Derrida1980} B. Derrida, Phys. Rev. Lett. 45 (1980) p.79.
\bibitem{Eshelby1954} J.D. Eshelby, J. Appl. Phys. 25 (1954) p.255.
\bibitem{Eshelby1957} J.D. Eshelby, Proc. Roy. Soc. A 241 (1957) p.376.
\bibitem{Martinez2001} L.M. Martinez and C.A. Angell, Nature 410 (2001) p.663.
\bibitem{Pelletier2002} J.M. Pelletier, B. Van de Moortele and I.R. Lu, Mater. Sci. Eng. A 336 (2002) p.190.
\bibitem{Wen2004} P. Wen, D.Q. Zhao, M.X. Pan and W.H. Wang, Appl. Phys. Lett. 84 (2004) p.2790.
\bibitem{Zhao2007} Z.F. Zhao, P. Wen, C.H. Shek and W.H. Wang, Phys. Rev. B 75 (2007) p.174201.
\bibitem{Hu2009} L.N. Hu and Y.Z. Yue, J. Phys. Chem. C 113 (2009) p.15001.
\bibitem{Lu2003} J. Lu, G. Ravichandran and W.L. Johnson, Acta Mater. 51 (2003) p.3429.
\bibitem{Vogel1921} H. Vogel, Phys. Zeit. 22 (1921) p.645.
\bibitem{Tammann1926} G. Tammann and W. Hesse, Z. Anorg. Allg. Chem. 156 (1926) p.245.
\bibitem{Fulcher1925} G.S. Fulcher, J. Am. Ceram. Soc. 8 (1925) p.339.
\bibitem{Fisher1928} R.A. Fisher and L.H.C. Tippett, Proc. Cambridge Phil. Soc. 24 (1928) p.180.
\bibitem{Gnedenko1943} B. Gnedenko, Ann. Math. 44 (1943) p.423.
\bibitem{Gumbel2004} E.J. Gumbel, Statistics of extremes (Dover, 2004)
\bibitem{Bouchaud1997} J.-P. Bouchaud and M. M\'{e}zard, J. Phys. A 30 (1997) p.7997.
\bibitem{Carpentier2001} D. Carpentier and P. Le Doussal, Phys. Rev. E 63 (2001) p.026110.
\bibitem{Fyodorov2008} Y.V. Fyodorov and J.-P. Bouchaud, J. Phys. A 41 (2008) p.372001.
\bibitem{Angell2000} C.A. Angell, K.L. Ngai, G.B. McKenna, P.F. McMillan and S.W. Martin, J. Appl. Phys. 88 (2000) p.3113.
\end{thebibliography}
\end{document}